# Molecule-by-molecule nanoresolved picosecond Purcell-enhanced emission at room-temperature in hollow plasmonic nanocone metamaterials

**R. Margoth Córdova-Castro, [1, 2 *] Dirk Jonker,[3,4]† Clement Cabriel,[1]† Mario Zapata-Herrera,[5] Bart van Dam,[1] Yannick De Wilde,[1] Robert W. Boyd, [2, 6] Arturo Susarrey-Arce,[3,4] Ignacio Izeddin [1 *] and Valentina Krachmalnicoff [1 *]**

[1] Institut Langevin, ESPCI Paris, Université PSL, CNRS, Paris, France.

[2] Department of Physics, University of Ottawa, Ottawa, ON, Canada.

[3] Department of Chemical Engineering, MESA+ Institute, University of Twente, Enschede, The Netherlands.

[4] Mesoscale Chemical Systems, MESA+ Institute, University of Twente, Enschede, The Netherlands.

[5] Donostia International Physics Center (DIPC), San Sebastian, Spain.

[6] University of Rochester, Institute of Optics, Rochester, NY, USA.

†These authors contributed equally.

*Corresponding authors: rcordova@uottawa.ca, ignacio.izeddin@espci.fr, valentina.krachmalnicoff@espci.fr

## Abstract

Nanoscale variations in the electromagnetic environment strongly influence quantum emission processes, yet direct quantitative visualization of these effects at the single-emitter level remains challenging. Here, we introduce a single-molecule fluorescence lifetime nanoscopy technique that quantitatively maps light–matter coupling within three-dimensional electromagnetic nanoenvironments under ambient conditions. By combining single-molecule super-resolution microscopy with time-correlated single-photon counting (TCSPC), we read the local electromagnetic environment of each molecule from its position and its excited-state lifetime. Using wafer-scale arrays of hollow plasmonic nanocones, we map position-dependent variations in the local density of optical states (LDOS) along the nanocone axis. We observe pronounced LDOS gradients that evidence a shortening of the fluorescence lifetime from nanoseconds to few picoseconds. The measured decay rates along the nanocone axis with maximum value at the top ring, yielding a total spontaneous-emission rate enhancement of more than two orders of magnitude at the single-molecule level, reaching an enhancement of the total spontaneous-emission rate of up to two orders of magnitude at the single-molecule level, in agreement with numerical simulations. These results provide direct experimental access to three-dimensional electromagnetic landscapes and establish a non-perturbative platform for quantitative molecular nanosensing in complex photonic architectures. The approach opens new opportunities for engineered room-temperature photon sources, vectorial field sensing, and scalable integrated nanophotonic technologies.

## Introduction

Significant advances in controlling light emission, absorption, and propagation through designed materials have enabled new physical phenomena and applications across both classical and quantum regimes, while simultaneously introducing challenges associated with light–matter interactions at nanometer length scales[1–3]. Precise control of single-photon emitters is a key requirement for emerging integrated quantum technologies[4,5] including secure communications[6,7], quantum computing[8], and ultrasensitive quantum sensing of electromagnetic fields[9–11]. Yet, realizing both efficient quantum light sources and nanoscale quantum sensors requires a quantitative understanding of light–matter interactions in complex electromagnetic environments, where nanometer-scale variations in the local density of optical states (LDOS) can profoundly influence spontaneous emission dynamics and therefore the performance of the single-photon source.

A powerful route to achieving such control is through engineering of the photonic environment surrounding an emitter. By tailoring the optical modes available for emission, nanophotonic structures can strongly modify radiative processes and thereby enhance single-photon-source efficiency [12,13]. The spontaneous emission rate is governed by the LDOS, which describes the number of optical modes available for radiation to an emitter at a given position and frequency. Variations in the LDOS alter emission decay dynamics and are commonly quantified by the Purcell factor, which expresses the enhancement of the photon emission rate relative to free space [12–19].

Achieving and understanding single-molecule control over spontaneous emission therefore requires simultaneous engineering of the electromagnetic environment and precise control and detection of emitter position and transition-dipole orientation [14–19]. Nanostructured media provide a versatile platform for this purpose by enabling strong field confinement and enhanced emitter–matter coupling. In particular, metallic nanostructures—including nanoparticles, antenna dimers, periodic arrays and plasmonic metamaterials—have emerged as powerful systems for manipulating electromagnetic nanoenvironments and accessing a broad range of optical phenomena [20–23]. Their localized surface plasmon resonances (LSPRs) confine light to deeply subwavelength volumes and can be tuned through geometry, size and composition. Such structures have enabled different applications such as sensing and substantial Purcell enhancement, often under cryogenic conditions, through spectral matching between plasmonic resonances and quantum-emitter excitation or emission frequencies [16–34].

The electromagnetic field confinement generated by conventional plasmonic nanostructures are highly localized at nanometric scale dimensions. While the rich modes available in nanostructures boost the single-emitter performance, despite many advances, challenges persist in quantitatively probing the enhanced emission of an isolated emitter coupled to the nanoantenna and observing it at nanometer resolution scales [35-40], mainly because this emitter-matter interaction occurs within electromagnetic modes confined in a diffraction-limited mode volume of tens of nanometres [39-53]. For major quantum applications that require precise control and detection of bright and fast single-photon emission, the key limitations to directly correlating the emitter position with the spontaneous emission dynamics include: (i) positioning emitters within regions of extreme field confinement and enhancement; (ii) interrogating individual emitters without perturbing the local environment; (iii) discriminating single-emitter signals from background fluorescence and neighboring emitters; and (iv) determining emitter positions with sufficient nanoscale precision to resolve local variations in the LDOS.

Here we present a strategy to obtain bright and fast molecular photon sources at room temperature and characterize one-by-one their enhanced light–matter interactions observed with nanoscale resolution. We employ molecular emitters as quantum probes because of their small size, well-defined transition dipole moments, and tunable emission spectra [28], properties that also make them attractive sensors of local pressure, strain, temperature and electromagnetic fields [9,10,35–40]. Using single-molecule lifetime nanoscopy, a non-invasive wide-field methodology, we probe enhanced light–matter interactions one emitter at a time, correlating fluorescence lifetimes with molecular positions at picosecond temporal resolution and nanometer localization precision. We combine a wafer-scale plasmonic material featuring nanocones arrayed on a lattice [41] to encode the nanoposition of the emitters in the LDOS variation with a custom-built far-field single-molecule fluorescence lifetime imaging microscope (smFLIM)[40,42–44]. This yields direct nanoscale mapping of the position-encoded LDOS variations of individual molecular single-emitters [45-47].

Our approach relies on a highly anisotropic platform, a dispersive hollow plasmonic metamaterial[41] that strongly modifies the local photonic environment, with control provided through excitation polarization. Compared with conventional nanorod geometries, conical nanostructures exhibit richer modal responses[52–55], smaller mode volumes and stronger intensity field enhancement at their apexes. Furthermore, hollow plasmonic architectures can provide tighter electromagnetic confinement while reducing dissipative losses owing to their reduced metal volume[56,57,58]. Our three-dimensional anisotropic geometry, an array of hollow plasmonic truncated nanocones, supports both transverse and longitudinal plasmonic modes. The excitation of these modes and the resulting localized hotspots are governed by the angle of out-of-plane p-polarized illumination. Under oblique illumination, simulations predicts and intensity field enhancement exceeding 600-fold at the top ring according to our simulations. These structures can be fabricated over centimeter-scale areas while maintaining nanometer-scale control over individual nanocone morphology[41]. Fully exploiting such plasmonic architectures, however, requires a quantitative understanding of their modal response and of single-emitter coupling within the three-dimensional geometry.

We chemically attached the molecular emitters uniformly to the gold surface of the nanostructures. Photoactivation was used to reach the single-molecule regime, so that the thousands of individual molecules could be detected one at a time and summed up to render a nanometre-scale position-encoded map of the emission modification. We show that the measured decay-rate-enhancement depends strongly on emitter position as well as on the angle of illumination, and, from numerical simulations, on the orientation of transition dipole. Together these reveal a highly heterogeneous light–matter interactions throughout the structure. The hollow conical architecture creates a strongly dispersive electromagnetic environment in which light–matter interaction is highly position-dependent, experimentally yielding a broad distribution of spontaneous emission lifetimes— spanning from nanoseconds to picoseconds —and up to 100× enhancement of spontaneous emission decay rates compared with the molecule on a glass substrate at room temperature.

Our work establishes a scalable framework for super-resolution quantum nanometrology by enabling non-invasive mapping of nanoscale electromagnetic environments through single-molecule observables. By combining wafer-scale hollow plasmonic metamaterials with nanometer-resolved lifetime imaging, we directly connect the local photonic structure to LDOS variations and spontaneous emission dynamics at the single-emitter level. Beyond demonstrating room-temperature Purcell enhancements approaching two orders of magnitude, this work is relevant to applications limited by spatial resolution, single-photon efficiency, and vectorial field mapping. This platform provides a general route for designing and characterizing next-generation scalable quantum photonic devices, high-efficiency single-photon sources and ultrasensitive optical sensors without the need for cryogenic operation.

## Results

We experimentally resolved and quantified how nanoscale LDOS variations in a three-dimensional plasmonic environment translate into extreme, position-dependent modifications of single-molecule spontaneous emission at room temperature.

### Angle-tuneable hollow plasmonic nanocones create extreme fields and mode dispersion in three-dimensional electromagnetic landscapes

In this section, we describe the plasmonic properties of the nanomaterial platform used to enhance single-emitter photophysics at room temperature.

Motivated by the flexibility of modern nanofabrication—where geometric parameters can be widely tuned to engineer plasmonic modes and tailor the photonic density of states—we leverage three-dimensional (3D) metallic nanostructures with axial anisotropy. This additional degree of freedom enables richer electromagnetic resonances, including longitudinal modes excited under oblique illumination[52–54]. While colloidal synthesis offers broad control over nanoparticle size and shape to tune the LSPR across the visible and near IR spectra range, conventional nanopatterning techniques such as electron-beam lithography (EBL) provide higher precision over dimensional parameters, particle placement, and array spacing for 2D nanostructures and metasurfaces. Self-assembly approaches have produced free-standing nanorod-[52,54] and nanocone-based[53] 3D metamaterials, demonstrating dispersive mode generation and strong field enhancement[52–55].

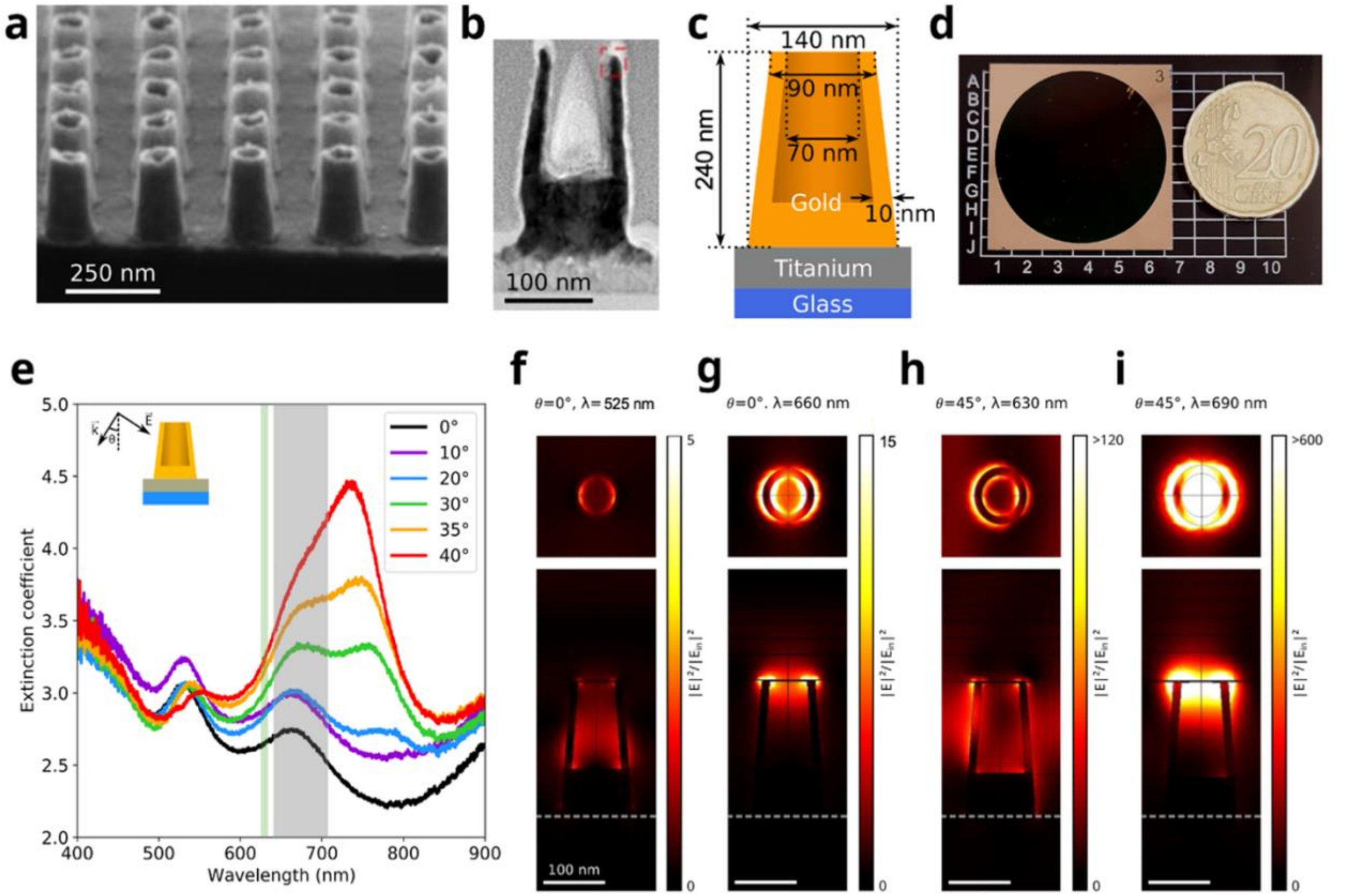

*Fig. 1: Engineered 3D electromagnetic environment (a-c) Images of the hollow gold truncated nanocones (HTC) fabricated on a square lattice with a period of 250 nm. (a) Scanning electron microscopy (SEM) image showing and the array of HTC. (b) Transmission electron microscopy (TEM) section of a single structure. (c) Schematic summarizing the dimensions of the unit cell containing a single HTC. (d) The HTC square lattice can be fabricated uniformly in a centimeter-scale chip, as shown in the photograph. (e-i) Optical characterization of the plasmonic resonances supported by the plasmonic nanostructure array. (e) Measured extinction spectra when the array is illuminated with a p-polarized plane wave at different illumination angles θ with respect to the axis of the nanocones. (f-,i) Calculated near-field intensity field enhancement maps for (f,g) normal and (h,i) oblique illuminations at different wavelengths. The top and bottom images display the top and side views, respectively. Scale bars: 100 nm.*

Among the various geometries for plasmonic light control, hollow nanostructures provide more confined electromagnetic fields, strong field enhancement, reduced dissipative losses, and reduced quenching of localized surface plasmon resonances through plasmon hybridization[57,58]. These features make hollow geometries particularly promising for designing plasmonic materials with superior performance compared with solid nanostructures. Owing to their intense plasmonic fields, hollow metal nanostructures have been explored for sensing, surface-enhanced Raman scattering (SERS), photothermal therapy, drug delivery, and catalysis, and can outperform their solid counterparts[41,57,58]. However, scaling fabrication while maintaining uniformity and reproducibility across large areas remains challenging, particularly for wafer-scale chip architectures. Here we combine hollow nanostructures with dispersive plasmonic modes in a hollow truncated nanocone square nanolattice which can be reliably fabricated up to to the $cm^2$ scale.

Hollow gold truncated nanocone (HTC) arrays were fabricated using a redeposition process that combines subtractive hybrid lithography (SHL) with Talbot lithography and i-line photolithography for precise nano- and macroscale patterning[41]. Additional fabrication details are provided in the **Methods** section. **Fig. 1a–c** shows the geometry of the HTCs and their compact dimensions (see also **Supplementary Fig. S1**). The bottom and top outer diameters are 140 nm and 90 nm, respectively, with an inner top diameter of 70 nm, resulting in a top-ring thickness of ~10 nm. The nanostructures are 240 nm high and arranged in a 250 nm-pitch square lattice (**Fig. 1a**). The material can be fabricated uniformly up to the centimeter scale (**Fig. 1d**).

The hollow truncated nanocones support a rich set of modes because of their longitudinal asymmetry. To characterize how the excitation field is modified over a broad spectral range, we measured the optical response under p-polarized white-light illumination while varying the incidence angle from normal to oblique. The structures display a strongly dispersive response with plasmonic modes across the visible spectrum. **Fig. 1e** shows the measured extinction spectra for the array immersed in water as the incidence angle θ increases from 0° to 40°. At angles smaller than 20°, two prominent extinction peaks appear near 525 nm and 660 nm. As θ increases, a third peak emerges near 780 nm around θ = 20° and strengthens at larger angles, indicating the excitation of a longitudinal mode.

Full-wave simulations using the Finite Element Method (FEM) were performed to analyze the plasmonic modes and their near-field distribution and enhancement under both normal and oblique incidence of p-polarized plane waves in an infinite square array of Au hollow truncated cones (**Fig. 1f-i**). The field enhancement maps were generated at wavelengths corresponding to the resonances observed in the calculated extinction spectra (**Supplementary Fig. S2**). The simulations reveal that, at normal illumination, the electromagnetic field is broadly distributed along the 3D structure, while at 45° incidence, the field is highly localized at the top of the truncated cone. The hotspot intensity localized in the top ring reaches values of extreme intensity field enhancement (>600). This behavior is attributed to the conical geometry's efficiency in exciting the longitudinal localized surface plasmon resonance (LSPR), resulting in extreme field enhancement at the top of the nanostructure at higher incidence angles—unlike nanorod structures, where the longitudinal dipolar LSPR is primarily confined at the base [53]. The tilted walls (relative to the vertical axis) promote strong field confinement at the apex, while the hollow architecture further concentrates the electromagnetic field within a very small volume around the top ring.

Simulations indicate that the field enhancement depends on the illumination conditions and can be tuned by increasing the excitation angle. Results for various wavelengths and angles are provided in **Supplementary Fig. S2**. This 3D electromagnetic landscape, together with the incident field, defines the excitation conditions experienced by single emitters coupled to these nanostructures and provides the basis for studying enhanced emission throughout the 3D nanoenvironment.

Despite remarkable progress in achieving plasmonic mode volumes that enhance spontaneous emission rates, obtaining a detailed, quantitative understanding of emitter–nanostructure interactions at nanometer length scales remains challenging[14–19,34,40,43,44]. Such understanding is essential for the rational design of quantum photonic devices and requires experimental methods capable of probing individual emitters within highly confined, spatially heterogeneous electromagnetic fields[19,34,40,43,44]. In the following section, we use far-field super-resolution microscopy to study emitter–plasmon coupling in these three-dimensional local fields at the single-molecule level with nanoscale spatial resolution[40,42–47].

## Nanoscopy of single-molecule emission on the plasmonic nanolattice reveal diverse molecular-plasmon coupling and picosecond lifetimes at room temperature

Fluorescence lifetime measurements provide direct experimental access to the LDOS at an emitter's position[14–18]. To investigate single-emitter emission coupled to our dispersive hollow plasmonic nanostructures, we chemically attached organic molecules to the gold surface of the hollow truncated nanocones. Unlike many other quantum emitters, organic molecules are effectively zero-dimensional probes with well-defined transition dipoles and nanometer-scale dimensions, making them particularly suitable for sampling confined optical environments[9,10,35,36]. Their minimal steric hindrance enables access to small features deep within complex samples and are compatible with three-dimensional mapping of electromagnetic landscapes. This contrasts with larger semiconductor quantum dots, whose dimensions and shape dispersion can limit dense nanoscale sampling[3,6,28].

We performed single-molecule fluorescence lifetime imaging (smFLIM), interrogating each molecule individually and localizing its position with nanoscale resolution. Inspired by single-molecule localization microscopy (SMLM) techniques such as photoactivated localization microscopy (PALM) and stochastic optical reconstruction microscopy (STORM), smFLIM enables super-resolution localization of individual fluorescent molecules with a spatial precision around 15 nm while simultaneously measuring their spontaneous emission lifetimes. To achieve this combined temporal and spatial resolution, we developed a custom-built far-field single-molecule super-resolution microscope. This system correlates signals from an electron-multiplying CCD (EMCCD) camera with those from an array of eight single-photon avalanche diodes (SPADs) integrated with a time-correlated single-photon counting (TCSPC) system, providing picosecond temporal resolution[40,42,43,44,45,46,47].

The principle of the experiment is illustrated in **Fig. 2a–b**. The sample is labeled with photoactivatable fluorophores, and the photoactivation illumination power is tuned to reach the single-molecule regime. Unlike approaches in which fluorophores are embedded in a polymer matrix—leading to a random 3D spatial distribution—our sample is chemically functionalized to form a thin monolayer of fluorophores that bind specifically to the gold surface at a distance of approximately 4 nm (see **Methods**). This enables the study of single-molecule fluorescence under polarized excitation at different incidence angles. Under wide-field nanosecond-pulsed excitation, emitted photons pass through a 50:50 beamsplitter and are detected by an EMCCD camera (to localize each molecule) and by a SPAD array (to measure the spontaneous emission lifetime). We acquired data continuously for ~5 h per measurement, cumulatively recording the response of the enhanced emission of each detected molecule to reach a high density of localized molecules (>5,000 molecules/μm$^2$), simultaneously recording the image in the EMCCD camera with exposure time of 30 ms per frame, while collecting time-tagged photon data detected by the SPAD array. The photoactivation power was optimized to satisfy two conditions: (1) Point Spread Functions (PSFs) on the EMCCD are sufficiently sparse to be localized individually (i.e., separated by more than the diffraction limit, ~250 nm), with each PSF corresponding to a single emitter; and (2) each SPAD detects signals from at most one molecule at a time. In practice, if two molecules emit simultaneously within the 10-μm-wide region associated with the SPAD array, their PSFs must be separated by more than 1 μm—the center-to-center distance between SPADs—otherwise the corresponding data is discarded. This approach thus requires an emitter density below 1 molecule per μm$^2$ per 30 ms time frame, which is critical for our technique to measure the emission properties of single emitters independently, avoiding simultaneous detection of neighboring molecules.

To determine the total spontaneous emission decay rate, Γ (the sum of radiative and non-radiative decay rates), for each single emitter, we used pulsed excitation in conjunction with time-correlated single-photon counting (TCSPC). TCSPC records a fluorescence decay histogram as a function of the delay between the excitation pulse and the detection of an emitted photon. The histogram follows an exponential decay, where $\tau = 1/\Gamma$ is the excited-state lifetime. The repetition rate was set to 39 MHz to allow sufficient time for relaxation to the ground state before the next pulse. Decay rates were obtained by fitting decay curves with a single-exponential function convoluted with the instrument response function (IRF). Events that do not exhibit mono-exponential decays (a key criterion for single emitters under our conditions) were discarded, as they may indicate overlap of two molecules.

The coordinates and fluorescence lifetimes of each detected molecule were then correlated to associate their spatial positions with their respective decay rates. Using this information, we accumulated single-molecule data to reconstruct a super-resolved Purcell enhancement map of single emitters with precise nanolocalization. Detection of thousands of molecules over the SPAD array field of view is essential to achieve sufficient sampling when performing the super-resolution imaging of the nanoscale lattice over an extended area with the superlocalized PSFs. Our smFLIM setup enables the measurement of lifetimes as short as a few tens of picoseconds and position precision of approximately 14 nm within a 10 µm x 1 µm field of view [40,44].

Compared with commercial SPAD cameras, the customized TCSPC system combined with the SPAD array offers an excellent balance of high detection efficiency (>30% at 650 nm) and low timing jitter (<60 ps FWHM), ensuring reliable temporal resolution across the extended sample area. Detailed information on sample preparation, experimental setup, and data processing procedures can be found in the **Methods** section.

Reconstructed super-resolved images are shown in **Fig. 2c–e**. These images reveal eight circular regions, each roughly 1 µm in diameter, corresponding to the areas in the camera's field of view conjugated with each SPAD in the array. The full detection density and intensity maps over the 10 µm² area, displayed in and 2 x10 µm, captured by the EMCCD camera, are presented in **Supplementary Fig. S3**. **Fig. 2c** displays the density map of detected molecules. This high spatial precision, around 15 nm, enables clear resolution of the 250 nm square lattice, with the measured sizes showing good agreement with the scanning electron microscopy images in **Fig. 2f**. Interestingly, while coupling between fluorescent emission and resonant structures has been reported to cause localization errors of several hundred nanometers due to PSF deformation[40,44], the accurate reconstruction of the lattice in our measurements suggests that this effect remains limited in the present study.

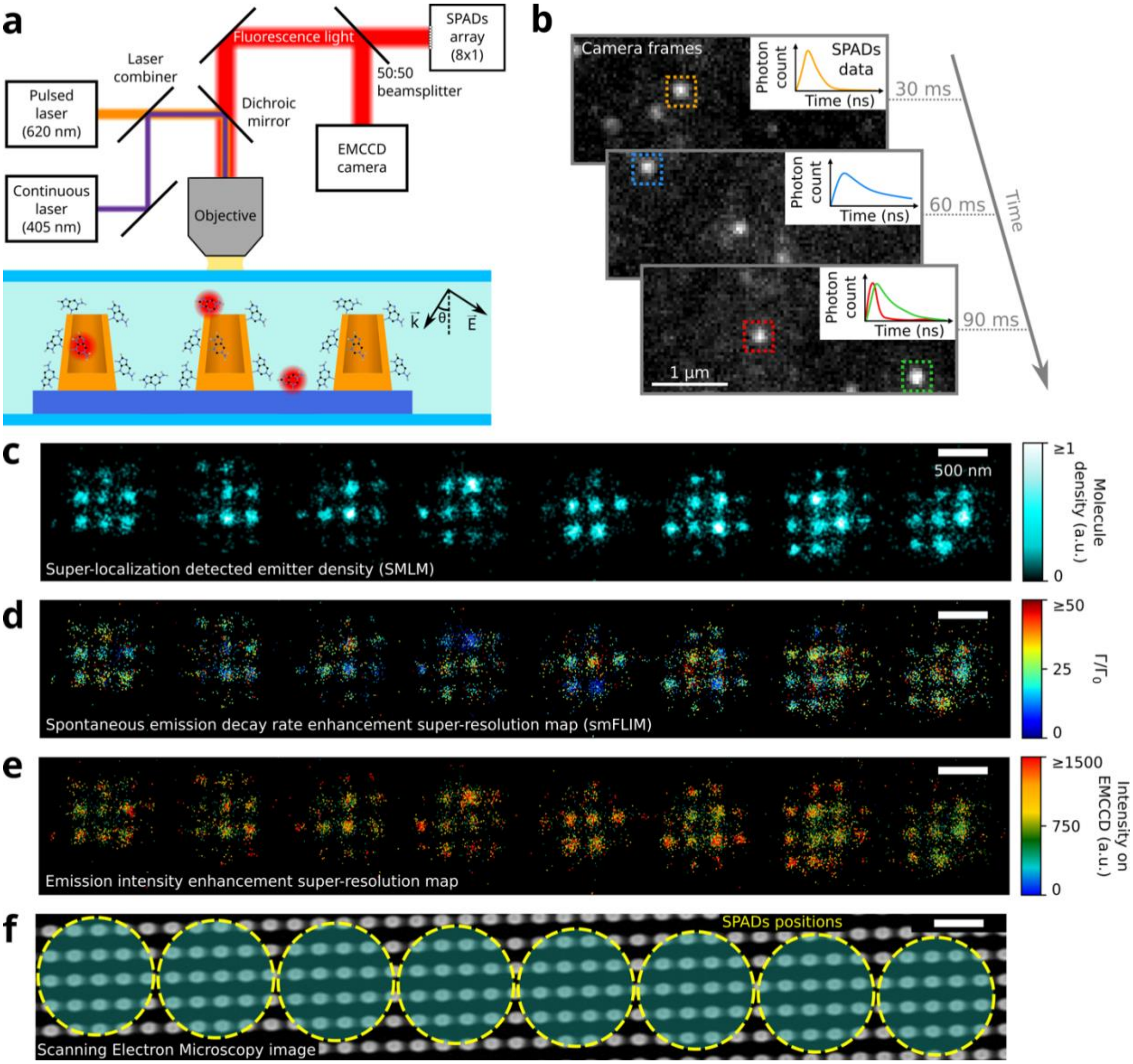

***Fig. 2: Nano-resolved enhanced emitter-matter interaction maps****. a) Schematic showing the optical setup and the sample mounting. b) Principle of the smFLIM experiment: on each EMCCD frame, the emitting molecules are detected (dotted color squares) and localized to measure their positions, which are correlated to the decay rate curves detected by the different SPADs (top right insets). This procedure allows the molecule-per-molecule mapping of the Purcell enhancement of the single emitters with a resolution below the diffraction limit. c-e) Super-resolved maps of single emitter detections in an area of 10 µm² for an oblique p-polarized pulsed laser illumination at 620 nm. Various types of information are displayed: the density of detected molecules, represented as a pixelized map (c), and the total decay rate enhancement (d) and collected fluorescence intensity (e) maps for each molecule. The latter two are displayed as scatter plots where each point is a detected molecule, the color standing for the decay rate enhancement or the intensity. f) SEM image showing the corresponding field of view and the positions of the eight SPADs (yellow dashed circles). Scale bars: 1 µm (b), 500 nm (c-f).*

**Fig. 2d** presents a super-resolved map of the total decay rate enhancement, normalized to the average value measured for the same fluorophore on a glass coverslip immersed in water ($\Gamma_0$ = 0.3 ns$^{-1}$). We observe pronounced spatial variations across the nanocones, consistent with position-dependent emitter coupling. **Fig. 2e** shows the corresponding super-resolved intensity map, defined as the integral of the Gaussian fit to each molecule's PSF, which similarly varies across the nanocones. Notably, the emission remains largely unquenched despite the <10 nm separation between the fluorophores and the metal. Together, these maps highlight the rich and heterogeneous light–matter interactions in this highly anisotropic hollow truncated plasmonic metamaterial.

## Super unit-cell reconstruction enables statistically robust single-molecule LDOS mapping

The strong variation in single-emitter properties upon coupling to the truncated plasmonic nanocones makes this approach well suited for nanoscale molecular sensing, including inference of emitter nano-position and orientation, as well as detection of electromagnetic variations within the 3D nanomaterial. To improve the statistical robustness of the analysis, we merged the data into a single unit cell by leveraging the square periodicity of the array and the symmetry of the nanostructures, as well as the reliability of the nanocone spacing. We thus superimposed thousands of detected events into a single *supercell*, as schematized in **Fig. 3a**. Briefly, the super-resolved maps in **Fig. 2c–e** were post-processed to identify the center of each nanostructure and to define a 2D square lattice. Although the lattice period (250 nm) is known with high precision, as we verified by overlaying SEM images with the detected lattice (**Supplementary Fig. S4**), the lattice origin and orientation were determined for each smFLIM acquisition. The single-molecule localization data were then partitioned into 250 × 250 nm unit cells centered on each HTC, and these unit cells were superimposed to form a single supercell containing all molecules detected within the eight-SPAD array.

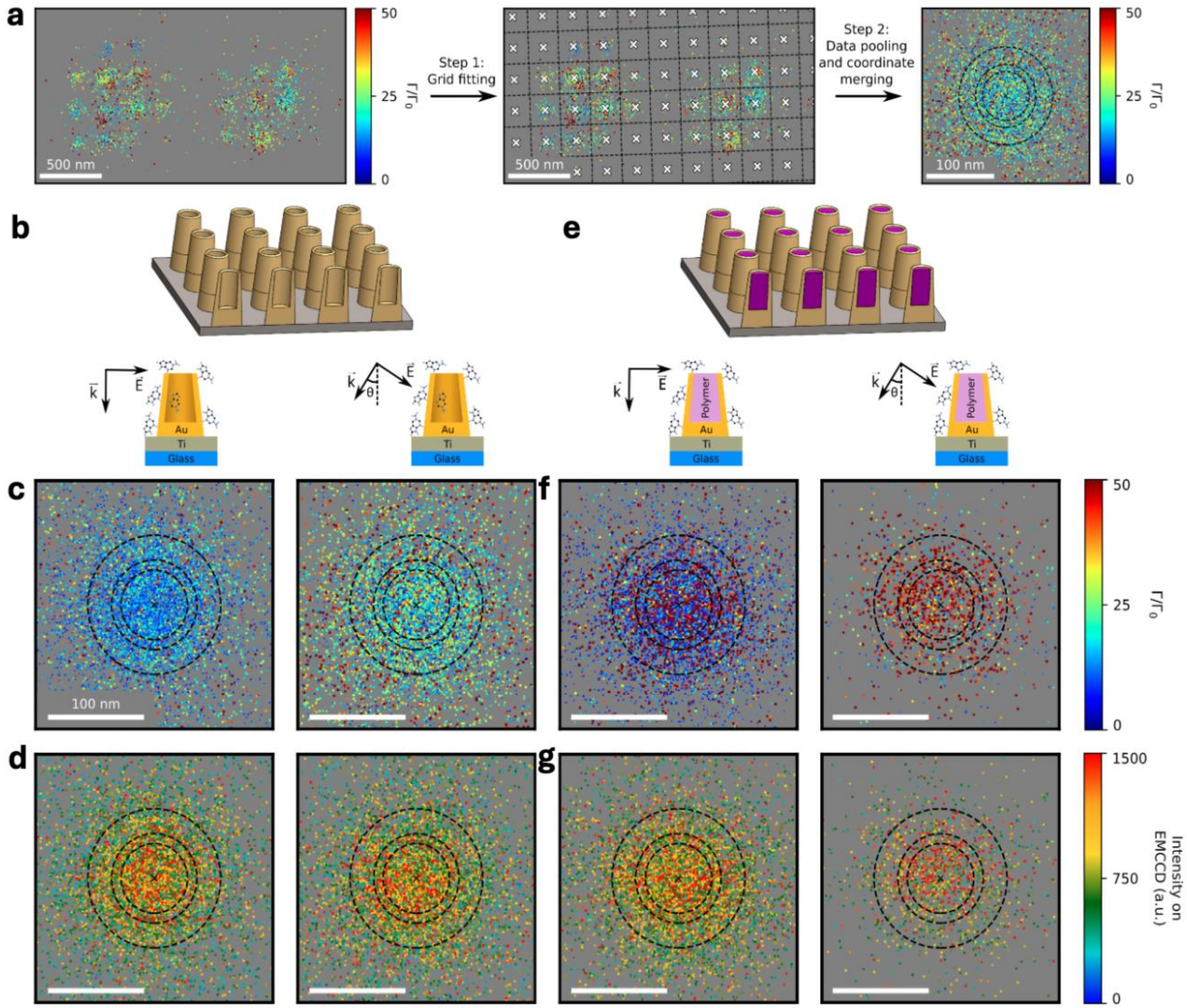

***Fig. 3: Spatial distributions of the measured decay rate enhancement on the supercell for empty and filled cavities, at normal and oblique illumination.*** *(a) Representation of the post-processing steps to merge the acquired data in the 10 µm² field of view (for clarity reasons, only a region of 2.5 µm², i.e. two SPADs, is displayed here) to a 250x250nm unit cell. (b) Schematic illustrating the normal and oblique illumination conditions for a sample where the cavities are empty. (c) Total decay rate enhancement maps in the 'supercell', normalized to the average value obtained for molecules attached on a glass coverslip* $\Gamma_0$ *=* $0.3\ ns^{-1}$*, for normal and oblique illuminations on a sample with empty cavities. (d) Intensity (i.e. number of photons detected on the EMCCD per 30-ms frame) maps in the 'supercell', for normal and oblique illuminations on a sample with empty cavities. The dotted circles stand for (from the outer to the inner one) the diameter of the base of the HTC, the outer diameter of the top ring and the inner diameter of the top ring (140 nm, 90 nm and 70 nm respectively), as defined in* ***Fig. 1c****. (e) Schematic illustrating the normal and oblique illumination conditions for a sample where the cavities are filled with polymer, preventing molecules from attaching inside the cavity. (f,g) Total decay rate enhancement (f) and intensity (g) maps in the 'supercell', for normal and oblique illuminations on a sample with polymer-filled cavities. All the maps are displayed as scatter plots, with each molecule being represented as one point. Scale bars: 500 nm (a left and center), 100 nm (a right, c,d,f,g).*

Since each SPAD has a diameter of approximately 1 µm—and thus contains roughly 8 complete hollow nanocones—the supercell contains approximately 64 times more molecules than a single nanocone. This significantly improves the statistical size and significance of the analysis (see **Supplementary Fig. S4** for more details). The results obtained with this supercell approach are summarized in **Fig. 3b–g**. The larger number of events enables a more detailed analysis of decay rate and intensity variations as a function of position within the unit cell, particularly with respect to the nanocone center.

**Fig. 3c** shows the decay rate enhancement measured within the supercell for molecules attached to the 3D nanostructure. Higher decay rates are observed under oblique illumination than under normal incidence, and in some cases the emission rate is accelerated by more than 50×. This trend is consistent with enhanced excitation at the top ring, which increases the probability of exciting molecules that couple efficiently to longitudinal modes of the truncated conical geometry. In addition to changes in the total decay rate, we observe a clear position dependence in the fluorescence intensity detected on the EMCCD (i.e., the number of photons collected per molecule), which increases when the emitter interacts strongly with the plasmonic system compared with its value far from the nanostructure (**Fig. 3d**). Full datasets of molecule detections across the EMCCD and SPAD arrays are provided in **Supplementary Fig. S5** for normal illumination.

To explore the influence of molecular position and orientation, we conducted experiments where we prevented molecules from attaching to the inner surfaces of the nanocone by filling the cavity with a polymer, as schematically illustrated in **Fig. 3e**. As expected, this reduced the overall density of detected molecules (**Fig. 3f**). Similarly to the case without the polymer, the decay rate enhancement increases for oblique illumination. When we prevent the molecules from being deposited inside the hollow cone, the effect of the tilted geometry seems to be accentuated. The same behavior is also evident in the fluorescence intensity maps shown in **Fig. 3g**. **Supplementary Fig. S6** provides SEM images confirming that the polymer coating is confined within the cavity throughout the large-area sample. The full-area maps of detected molecules for both normal and oblique illuminations, with the cavity filled with polymer, are shown in **Supplementary Fig. S7** and **S8**, respectively.

Collectively, these observations suggest that the photophysical properties of the single molecules are optimized at the center of the hollow structure. This enhancement results from the interaction of the emitters with localized plasmonic hotspots, which strongly influence emission rates and intensities [24-31].

## Molecule counting and super-resolved nanoposition sensing via monotonical enhancement of spontaneous emission measurements

To explore how the spontaneous emission decay rate depends on the spatial position within the nanostructure, we examined the normalized decay rate as a function of the emitter's distance from the nanocone center. The strength of our technique lies in simultaneously recording the emission properties and positions of individual emitters, enabling detailed correlations and analyses at the molecular scale. **Fig. 4a–d** present a point cloud visualization of the normalized decay rate in the (x, y, Γ) space, merging data from thousands of emitters within a supercell across the four different experimental configurations. To better understand the spatial variation, **Fig. 4e–h** depicts the radial distribution of the same data, segmented into distinct decay enhancement intervals. The black dashed lines mark the radii corresponding to the inner top radius (35 nm), outer top radius (45 nm), and radius of the base of the cone (70 nm).

We first consider the unfilled structures, where emitters can bind anywhere on the gold surface, including inside the cavity. Under normal incidence (**Fig. 4a,e**), the decay rate enhancement is mostly below 20. In contrast, under oblique illumination (**Fig. 4b,f**), the decay rate increases substantially, with many values exceeding the maximum temporal resolution of our setup ($\Gamma_{max} = 50\ ns^{-1}$). Most molecules localize near the supercell center, and the broad range of measured decay rates reflects the strong sensitivity of the measurement to the emitter's precise position within the nanostructure.

As shown in **Fig. 4g,h**, filling the cavity with a polymer results in a pronounced decrease in the occurrence of decay-enhancement values between 20 and 30, consistent with the polymer preventing molecules from binding inside the cavity, where moderate enhancement is expected. Under normal illumination with the cavity filled, most decay enhancements remain below 20, with only a small fraction exceeding 40.

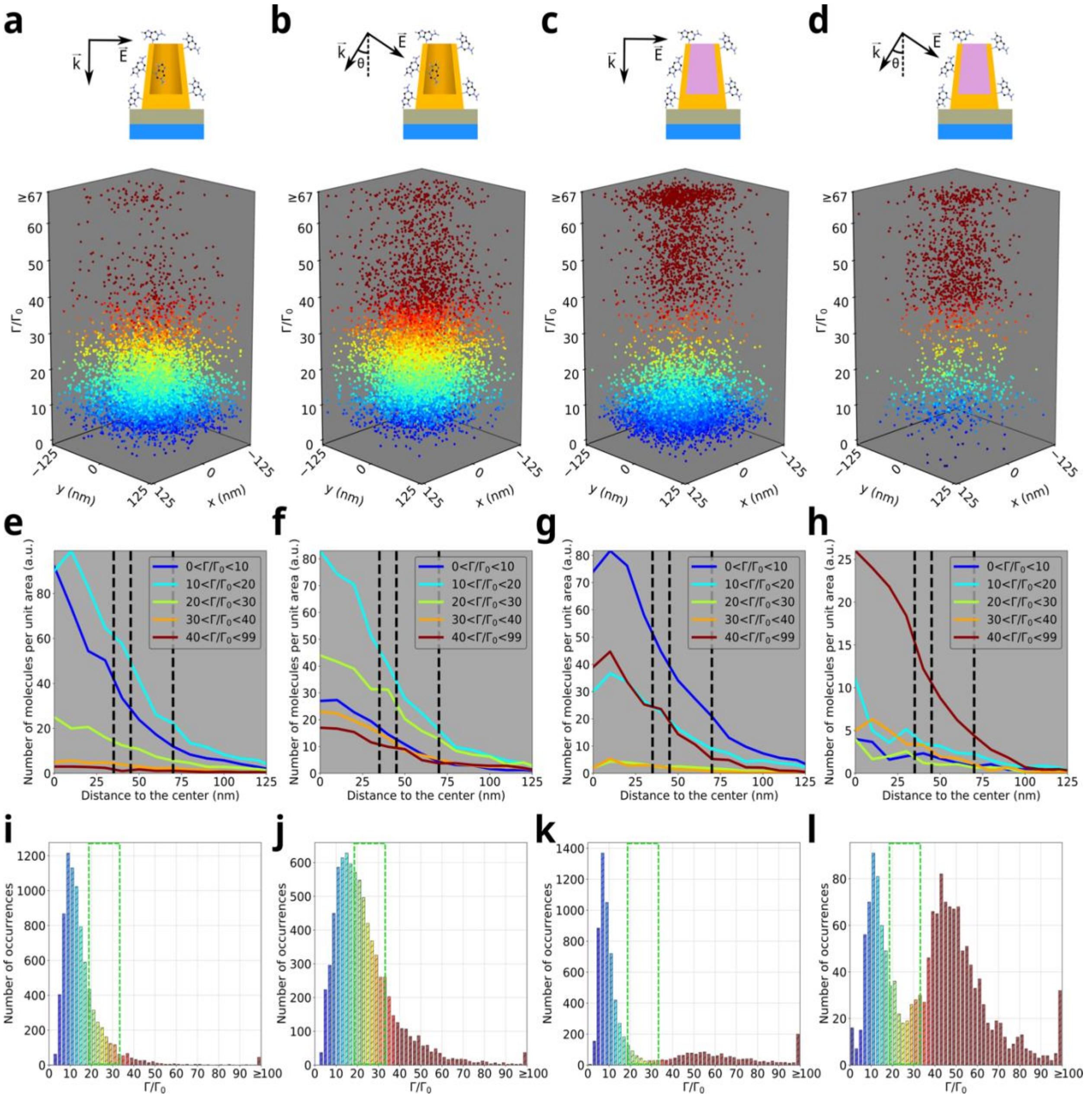

***Fig. 4: Molecule nanoposition sensing with spontaneous emission rate enhancement and quantification of Purcell enhanced single-molecules*** *a-d) Spatial distributions of the measured decay rate enhancement of all the data, plotted as a supercell for different experimental conditions. Each dot represents a molecule, and the color and height encode its measured decay rate. e-h) Radial distributions of the molecular density for different intervals of decay rate enhancement, plotted from the center to the edge of the supercell (125 nm). The black dashed lines stand for (from left to right) the inner radius of the top ring, the outer radius of the top ring and the radius of the base of the HTC (35 nm, 45 nm and 70 nm respectively), as defined in* ***Fig. 1c****. i-l) Histograms of the measured decay rate enhancement. The green dotted rectangles indicate the typical decay rate enhancement range for molecules inside the cavity, i.e. those that are blocked by the polymer. Throughout the figure, the results are presented for empty (a,b,e,f,i,j) and filled (c,d,g,h,k,l) cavities, and for normal (a,c,e,g,i,k) and oblique (b,d,f,h,j,l) illuminations.*

In contrast, under oblique illumination, the fraction of emitters exhibiting strong decay rate enhancement increases dramatically (**Fig. 4f,h**). This aligns with the understanding that these high enhancement instances ( > 40) originate from molecules located at the top of the hollow nanocones that are more efficiently excited under oblique illumination. As demonstrated by the simulations in **Fig. 1f-i**, oblique illumination excites a longitudinal mode of the hollow truncated cone, resulting in intense field enhancement at the structure's apex.

The distributions of decay rate enhancement across the four configurations (Fig. 4i–l) further support this interpretation. Specifically, filling the cavity with polymer reduces the occurrence of intermediate enhancements between 20 and 30 (compare Fig. 4k,l with Fig. 4i,j), consistent with suppressing contributions from emitters inside the cavity, whereas oblique illumination increases the fraction of strongly enhanced events (>40; compare Fig. 4l with Fig. 4k). These results indicate that lifetime measurements alone can provide a sensitive readout for molecular sensing of the 3D electromagnetic landscape within and around the nanostructure.

## Molecule-by-molecule measured fluorescence intensity and lifetime correlations provide signatures of radiative and non-radiative coupling regimes

A nanoantenna coupled to a single emitter can enhance the emitted power of an excited two-level system by modifying the electromagnetic impedance and the local density of optical states[14–17,22,25,31]. As shown in **Fig. 3**, our plasmonic system modifies both the total spontaneous emission rate and the detected fluorescence intensity. Fluorescence intensity reflects the combined effects of excitation enhancement, radiative rate enhancement, collection efficiency, and antenna-mediated radiation[25,27,29,30]. By contrast, lifetime measurements probe the total decay rate, including radiative and non-radiative contributions[14–17,30,34].

The spatial distribution of fluorescence intensity within the supercell is shown in **Fig. 5a–d**. Notably, emission is not quenched, even though the molecules are separated by less than 10 nm from the gold surface;

a radiative contribution persists in all cases. The highest intensities are observed near the supercell center, with lower-intensity events toward the edges. The intensity point cloud shows a smooth increase from the edges toward the center and a sharply peaked distribution at the supercell center, reflecting the super-localization precision of our method. This spatial pattern is consistent with the conical geometry of the nanostructure and the distance-dependent enhancement of both excitation and coupled emission. Importantly, the most strongly enhanced emission is observed at the supercell center, where the largest field enhancement is also predicted by simulations (**Fig. 1f–i**), confirming the critical role of geometry in dictating electromagnetic coupling at the single-molecule level.

Our measurements reveal a complex light–matter interaction involving both radiative and non-radiative channels. **Fig. 5e–h** illustrate this by showing the molecule-by-molecule correlation between the total decay rate and the detected fluorescence intensity. Interestingly, a bimodal distribution appears only when emitters are excluded from the nanocone cavity (**Fig. 5g,h**). This stochastic far-field approach enables simultaneous retrieval of the excited state lifetime and nanoscale position of individual molecular emitters, providing molecule-specific access to both far-field and near-field interaction regimes. Interpreting these results (**Figs. 4 and 5**) requires considering how the excitation field distribution varies with illumination angle (**Fig. 1f–i**). At normal incidence, the excitation predominantly enhances the region near the base, leading to an under-representation of molecules on the top ring, where the largest decay rate enhancements are expected. In contrast, under oblique illumination, the excitation is strongly enhanced near the top ring, which likely explains why higher decay rate enhancements are observed (**Fig. 4j,l**) compared with normal incidence (**Fig. 4i,k**), independent of whether the cavities are filled with polymer.

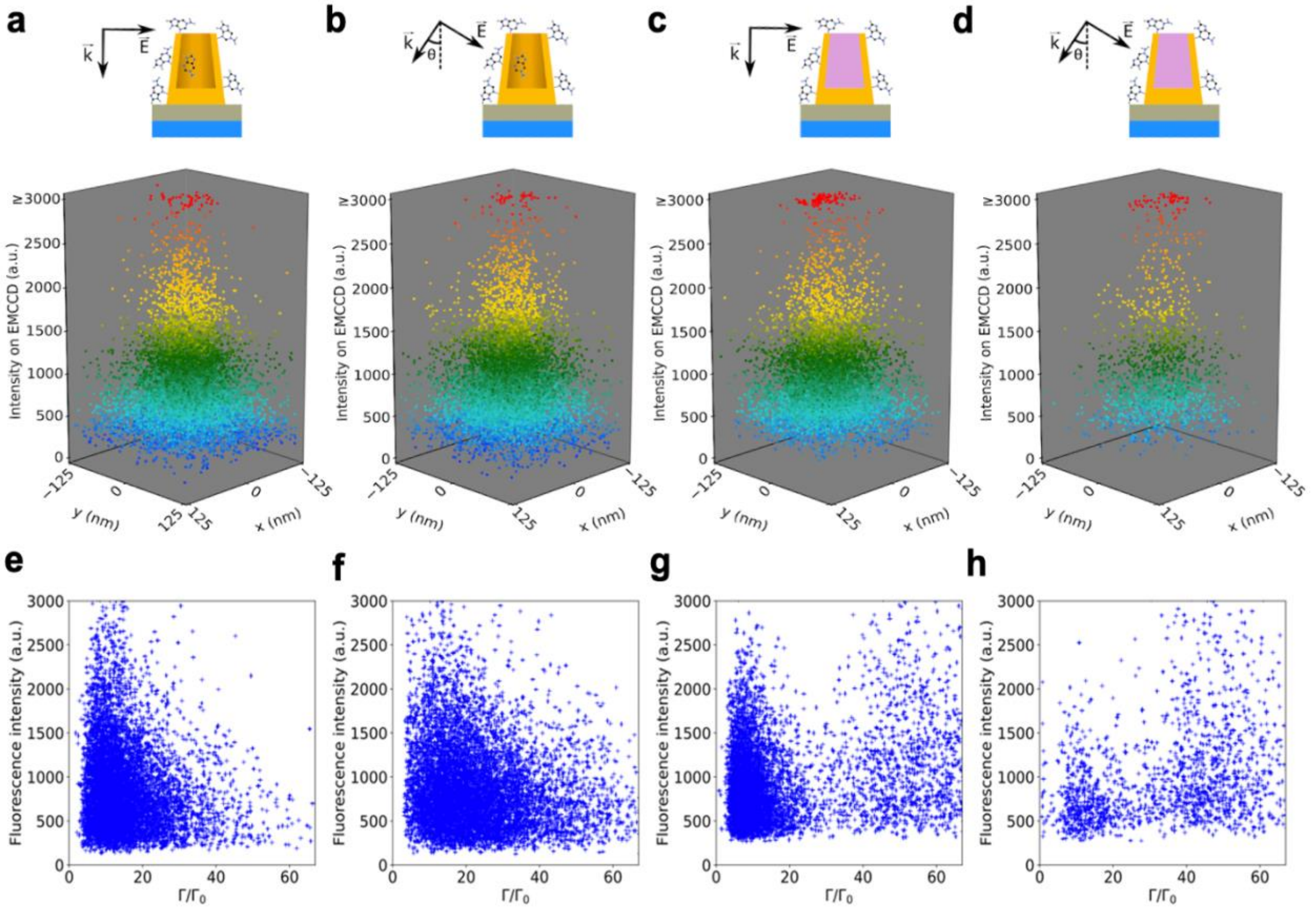

***Fig. 5. Measurement of the emission fluorescence intensity, highlighting the radiative contribution of the enhanced molecule-per-molecule light-matter interaction for single emitters coupled to the 3D hollow nanocones**. a-d) Spatial distributions of the measured fluorescence intensity (defined as the number of photons detected in the PSF per EMCCD frame, i.e. per 30 ms) for each molecule, displayed as a supercell scatter plot, each dot representing one detected molecule. The color and height stand for the intensity. e-h) Molecule-per-molecule correlation of the fluorescence intensity with the decay rate enhancement, displayed as a scatter plot where each cross represents one detected molecule. The total decay rate is normalized with respect to the measured value of the emitter decay rate on glass surrounded by water ($\Gamma_0$ = 0.3 $ns^{-1}$). Note that the positions of the molecules are not displayed in this representation. The data is presented without (a,b,e,f) and with the cavity filled (c,d,g,h) with a polymer, and for normal (a,c,e,g) and oblique (b,d,f,h) illuminations.*

## Point source electromagnetic simulations link position- and orientation-dependent LDOS to lifetime distributions

The LDOS—and therefore the decay rate enhancement—depends on both the emitter position relative to the nanocones and the orientation of its transition dipole moment[14–17,19,34]. To validate the experimental observations, we performed numerical simulations of the LDOS for multiple emitter positions and dipole orientations.

In the electric dipole approximation, appropriate for small quantum emitters at optical frequencies, the two-energy level system that describes the emission of a single photon in quantum mechanics can be seen as classical oscillating electric dipole with its characteristic dipolar moment orientation related with the transition from the exited state to the ground state, and higher-order oscillations can be neglected. In the classical scenario, the Purcell factor represents a change in the dielectric environment's impedance, linking the emitter's spontaneous decay rate to the impedance seen by a classical dipole antenna[16,17]. Similarly than for spontaneous emission rate, for a classical dipole, the total emitted power $P$ can be partitioned into radiative and non-radiative components. The radiative power, $P_r$, corresponds to the energy emitted into the far field, while the non-radiative power $P_{nr}$ accounts for absorption within the nano-environment, including dark and guided modes in the near field.

We numerically solved Maxwell's equations to analyze the interaction of single electric dipoles with the hollow truncated cones lattice at various positions $\boldsymbol{r}_i$, considering two distinct dipole orientations (in-plane and out-of-plane), across the visible spectrum. To assess the enhanced radiated power, we simulated finite square arrays of increasing size, monitoring the convergence of the results. Ultimately, a 10×10 lattice was used, which effectively approximates an infinite array for a dipole positioned within the first four HTC unit cells at the center.

The LDOS modification was then quantified as the ratio of the power emitted by a dipole—placed 4 nm from the gold surface with a specific orientation—to the power emitted by an identical dipole located on a glass substrate in water. This ratio provides access to the total LDOS modification, incorporating both near-field and far-field contributions. The results of these simulations are shown in **Fig. 6a-d**, for various positions and dipole orientations $P_x$, $P_z$, as depicted in the insets.

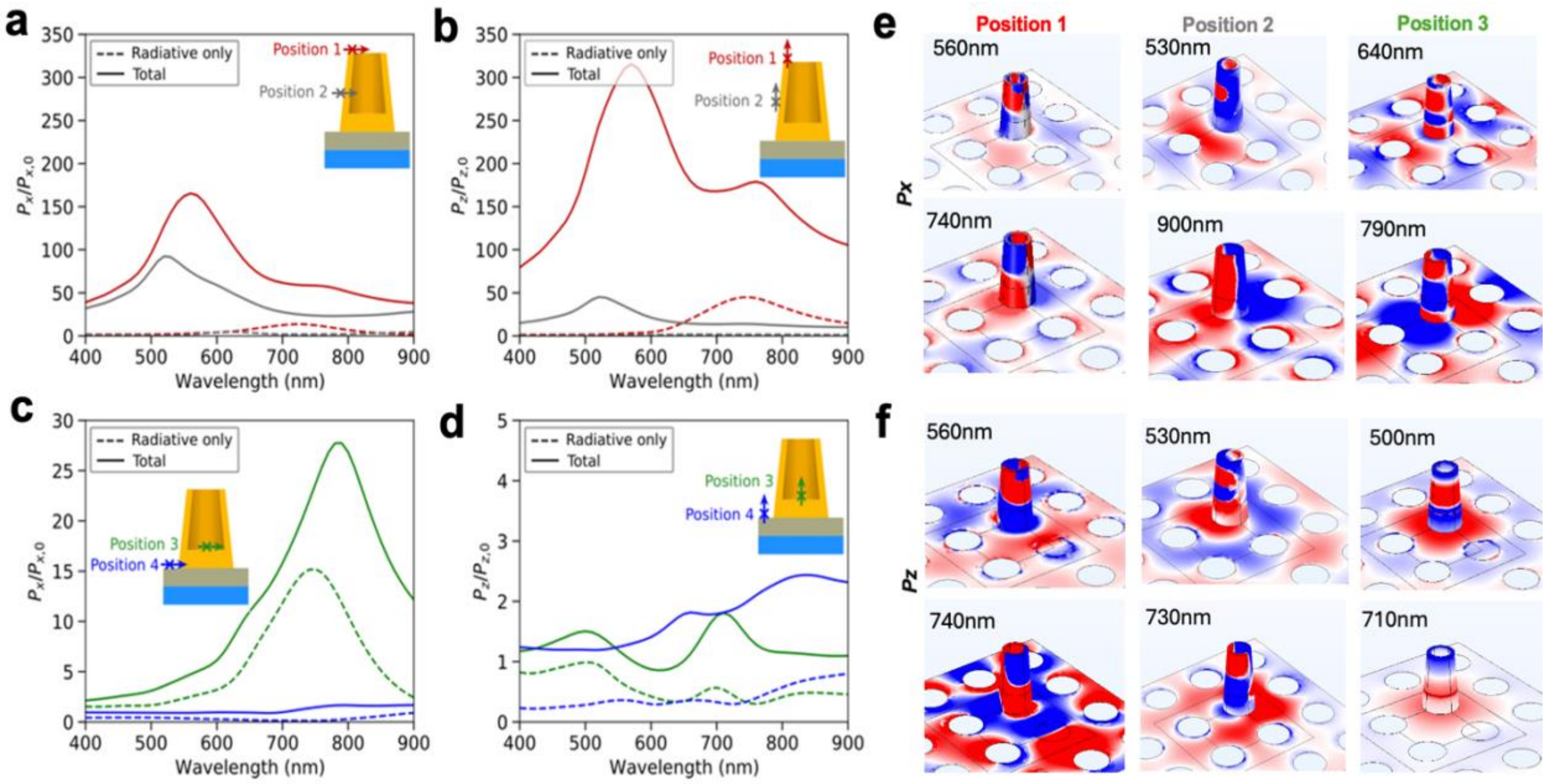

***Fig. 6: Calculated response of a point source coupled to the HTC.*** *a-d) Numerical calculations of total and radiative decay rate enhancements for a dipole attached to the outer wall of the HTC (separated by 4 nm) at different positions and with different dipole orientations (see the insets). The decay rate calculated on the top ring (position 1) is higher than the one sensed by the dipole on the outer wall (position 2) or inside the cavity (position 3) for both dipole moment orientations. Note the different y axes in (a-d). e-f) Charge distribution for electric dipoles at different positions, with different electric dipole polarizations and at different wavelengths. Each disk represents one HTC.*

The total enhanced decay rate can be computed by integrating the near-field Poynting flux over a spherical surface with a radius of a few nanometers enclosing the emitter, thereby capturing both radiative and non-radiative contributions. The calculated total decay rate enhancement varies strongly with emitter position and dipole orientation, with the highest values observed for dipoles placed on the top ring of the truncated cone (position 1), relatively high values for dipoles on the outer wall (position 2), intermediate values for dipoles inside the cavity (position 3), and the lowest values for dipoles near the base (position 4).

When no emitters are present inside the hollow nanostructure—such as when the HTC cavity is filled with polymer (i.e., no molecules at position 3)—we expect the measured enhancement distribution to be bimodal, with low values (<10 at position 4) and high values (>20 at positions 1 and 2). This is consistent with our experimental data, where a clear bimodal distribution is visible in **Fig. 4k,l** when no molecules are inside the cavity. The relative weight of the two distributions in the enhanced emission molecular counting is set not only by the local Purcell factor but also by the surface area of the nanoparticle, decreasing along the cone axis. At the top of the hollow cone, the ring is only about 10 nm wide, allowing a small fraction of the molecules only to deposit on the surface where according to our calculations the highest enhanced field. On the contrary, the outer wall and the base of the cone have a larger accessible metallic surface area where molecules can bind.

We performed additional calculations for positions 1, 2, and 4 with the cavity filled with a polymer of refractive index 1.6 and surrounded by water. Despite the contrast in refractive index between the water and the polymer, the Purcell factor calculations with the cavity filled change only slightly (**Fig. S9**). More precisely, we find similar total and radiative decay rates compared to the results in **Fig. 6a-d**, with only few nanometer shifts toward longer wavelengths at the maximum values in the spectra curve of LDOS. The simulations reveal a dominance of the non-radiative component of the LDOS in all scenarios; however, in some cases, there is a significant enhancement of the radiative decay rate. For instance, at position 1 around 750 nm, the radiative contribution is nearly enhanced 50-fold (**Fig. 6b**).

The wide range of decay rate enhancement values observed in our measurements can be better understood by examining the charge distributions (**Fig. 6e, f**) in the plasmonic metamaterial. At each position and dipole orientation, a variety of photonic modes are excited by the point source, leading to diverse electromagnetic responses. These calculated charge distributions vary across the spectral range, exhibiting dipolar, quadrupolar, octupolar, and more complex charge patterns depending on position and orientation. The conical geometry influences the emergence of different modes around both the top ring and the base, favoring the presence of dark modes within the visible spectrum. This is evident in **Fig. 6a** and **6b** where the maximum total enhancement does not coincide with the peak of the radiative contribution.

Localized plasmonic modes near the top ring can be seen in the charge distributions for position 1 (top of the cone), as shown in **Fig. 6e** and **6f**, for both dipole orientations. For the dipole at position 2, an enhanced LDOS peak appears around 530 nm, characterized by a hexapolar-like charge distribution for both polarizations. At longer wavelengths, in-plane polarization exhibits a dipolar charge pattern, while along the longitudinal axis, a quadrupolar-like distribution appears near 730 nm. These are qualitative examples of the possible modes supported by the structure. Additional simulations for dipole orientations along the y axis are presented in **Supplementary Fig. S10**, along with charge distributions at position 4, which display an even broader variety of modes. The pronounced position- and orientation-dependent behavior in this hollow noncanonical geometry explains the significant variation observed in our decay rate enhancements measurements.

## Discussion

Single-molecule nanophotonics enables the investigation of light–matter interactions at the emitter level with nanometer-scale resolution[19,29,35,36,40,43,44]. In this context, our single-molecule lifetime nanoscopy provides a non-invasive route to translate nanoscale changes in electromagnetic conditions into robust, position-encoded single-molecule observables. Our methodology, which temporally and spatially correlates enhanced single-emitter properties with super-resolved precision, opens opportunities for technologies limited by spatial resolution, single-photon efficiency and sensitive detection. Such as low-density sensing and vectorial field mapping[4–11,40,42–46].

Compared with smFLIM, many existing approaches do not provide access to the density of states deep within a 3D volume. In particular, most super-resolution studies of light–matter interactions rely on AFM-based scanning probes (e.g., scanning near-field optical microscopy, SNOM) and therefore primarily report near-surface information. By contrast, our methodology offers a minimally invasive route to study light–matter interactions not limited to the surface and without perturbing the system with external metallic scanning probes[18,19].

The structure studied here is a large-area array of three-dimensional hollow truncated nanocones arranged in a square lattice and fabricated uniformly at centimeter scale using a wafer-scale redeposition strategy[41].

This hollow plasmonic platform supports diverse, tunable light–matter interactions. The hollow conical geometry creates an anisotropic electromagnetic environment that is highly sensitive to molecular nanopositioning, producing strong spatial variations in the LDOS throughout the structure[52–55]. Compared with solid nanostructures, hollow architectures provide an extended plasmonic surface, strong field confinement, and tunable hybridized resonances while reducing the metal volume[57,58]. Under polarized plane-wave excitation, the tilted geometry supports a rich set of modes whose spatial profiles depend strongly on emitter position and illumination angle[52–55]—an effect that becomes especially clear when the cavity is blocked and emitters couple only to the outer surface. More broadly, our conformal functionalization approach reveals LDOS variations across the metamaterial thickness, suggesting opportunities for three-dimensional quantum sensing[9–11]. With alternative emitter-deposition strategies, the platform could position molecules or quantum dots at selected locations to achieve targeted spontaneous emission lifetimes[3,6,19,28]. Finally, the strongly structured local fields may provide a route to control and study chemical reactions as a function of electromagnetic intensity at nanometer and single-molecule resolution.

## Conclusions:

In this work, we introduce a scalable hollow plasmonic platform with rich optical modes and angle-dependent enhanced field, together with a single-molecule lifetime-nanoscopy methodology that probes enhanced molecular emission with nanometer-scale resolution. By combining super-resolution localization with time-resolved detection, we quantify position- and orientation-dependent LDOS variations within a 3D electromagnetic nanoenvironment and resolve nanosecond-to-picosecond fluorescence lifetime maps well beyond the diffraction limit.

We show that emission performance can be substantially improved by enhancing decay rates of individual molecular emitters, with the magnitude of the modification set by the emitter nano-position within the 3D hollow nanostructure and by the excitation angle. The hollow plasmonic conical geometry mitigates plasmonic losses while supporting a rich set of dispersive 3D modes, producing strong LDOS gradients and pronounced sensitivity to dipole orientation. By analyzing thousands of single-molecule lifetimes with high sensitivity and dynamic range, we map extreme variations in decay rate and fluorescence intensity as a function of molecular position, enabling the detection of subtle electromagnetic-field changes both at the surface and deep within the 3D near-field landscape. These super-resolved measurements directly link quantum light–matter interactions to the local photonic environment, demonstrating enhanced total decay rates and picosecond-scale spontaneous emission lifetimes at room temperature. Our sensing approach does not require invasive scanning probes. It collects and quantifies information about how the emission properties of bright, fast molecular photon sources vary with nanometre position, turning minute changes in emitter position and local environment into measurable single-molecule observables through the enhanced fluorescence intensity and the position-encoded lifetime. Integrating single-molecule detection with quantum sensors could also benefit chemical separation and analysis platforms. For example, combining quantum-sensing sensitivity with microfluidics may enable miniaturized devices capable of molecular-scale chemical separation and analysis[59,60]. Together, wafer-scale manufacturability and non-averaged Purcell-factor measurements provide a route toward reproducible, high-precision quantum sensing and scalable on-chip nanophotonic and quantum technologies.

# Methods

## Hollow truncated nanocone (HTC) array fabrication.

The fabrication of the hollow truncated nanocone (HTC) array uses subtractive hybrid lithography (SHL) combined with displacement Talbot lithography (DTL). DTL generates a self-image by exposing a periodic phase-shift mask with monochromatic illumination.

The large-area nanofabrication approach for precise nano- and macroscale patterning is described in Ref. [41]. The combined DTL/SHL process is advantageous because it offers high resolution, rapid patterning, flexibility in pattern design, cost-effectiveness, and a single mask layer. First, a resist pattern is generated on the substrate, enabling high-resolution control over nanoscale dimensions. Our method uses a single photoresist and two exposure steps to pattern the photoresist and bottom anti-reflective coating (BARC), forming a nanocone-like template. Gold is then deposited onto the patterned resist by sputter redeposition, forming dense arrays of hollow truncated nanocones arranged in a square lattice with a 250 nm pitch. The structures serve as a template for Au sidewall redeposition during angled ion beam etching (IBE). An angled IBE step controls the redeposition rate, in contrast to normal-incidence IBE reported previously.

The fabricated Au hollow truncated nanocones (HTCs) feature a solid base with a thickness of 80 ± 2 nm and a diameter of 140 ± 4 nm. The nanorims are 160 ± 10 nm high with a sidewall thickness of 12 ± 3 nm. After sputter redeposition, BARC remains inside the hollow cavity; these samples were used as “blocked-cavity” controls. To open the cavity, the BARC layer was removed using oxygen ($O_2$) plasma. Stripping the BARC/photoresist yields an open cavity and uniformly distributed HTCs at the wafer scale.

## SEM images

Following optical measurements, the sample was carefully sectioned using a wire cutter, intersecting the device area. Depending on the inspection needs, the resulting fragments were mounted on a stainless-steel sample carrier in either an upright or flat orientation. These mounted fragments were then transferred to a High-Resolution Scanning Electron Microscope (HR-SEM, Merlin, Zeiss) for detailed examination. This allowed both cross-sectional and top-view inspections to assess the structural features and surface properties of the fabricated nanocones.

## TEM images

A substrate with HTCs was coated with a 20 nm carbon film via sputter deposition and transferred to a high-vacuum chamber, where a ∼3 µm platinum layer was deposited using $Ga^+$-assisted chemical vapor deposition. A ∼1 µm thick, 15 µm long section was then cut using a 30 kV $Ga^+$ ion beam at a 4.8° angle to the square unit cell's axis. Before detaching the sample, a wire was welded to the Pt layer for handling. The sample was transferred to a TEM grid, the wire was milled off, and the sample was thinned to 50 ± 10 nm. The angled cut intersected an HTC every 12 periods, providing a representative cross-section.

## Optical characterization

Transmission was measured by mounting the sample on a rotation stage and illuminating it with p-polarized white light (Thorlabs). A linear polarizer–analyzer pair defined and verified the polarization state. The transmitted light was collected into an optical fiber and recorded with a UV–Vis spectrometer (Ocean Optics).

## Molecule-functionalized sample preparation and single-molecule regime optimization

After fabrication, the sample was labeled by incubating it for 30 min in a 40:5:1 mixture of phosphate-buffered saline (PBS), sodium bicarbonate solution (7.5% v/v, Sigma-Aldrich), and photoactivatable Abberior CAGE 635 (biotin conjugate) dissolved in dimethyl sulfoxide (DMSO, Sigma-Aldrich) at 0.6 mg $mL^{-1}$. Polystyrene fluorescent beads dispersed in PBS (Crimson FluoSpheres, 200 nm, Thermo Fisher Scientific) were added at low concentration as fiducial markers. The sample was rinsed with PBS between each step, and after the final step it was rinsed with demineralized water.

We recorded continuous acquisitions with both the EMCCD (30 ms exposure time) and the SPADs, and we optimized the photoactivation power to operate in the single-molecule regime. Specifically, we enforced two criteria: (i) PSFs on the EMCCD were sufficiently sparse to be localized individually (separated by more than the diffraction limit, ~250 nm), with each PSF corresponding to a single emitter; and (ii) each SPAD collected photons from at most one molecule at any given time (if two molecules emitted simultaneously within the ~10 μm-wide region associated with the SPAD array, their PSFs had to be separated by more than 1 μm—the SPAD center-to-center spacing—otherwise both events were discarded). These criteria were satisfied for an emitter density below 1 molecule $\mu m^{-2}$, which we achieved in the 3D structure by tuning the photoactivation laser power. Each molecule was thus interrogated from activation to photobleaching, and the activation rate was adjusted to approximately balance the photobleaching rate, maintaining a quasi-constant emitter density while satisfying the sparsity criterion. For long acquisitions (several hours), the HTC sample was mounted in a microchamber to stabilize the water volume and improve photophysical stability. We further optimized the signal-to-noise ratio by filtering the gold photoluminescence using an emission filter.

## Single-emitter fluorescence lifetime experiments

The optical setup is presented in Fig. 2a. smFLIM measurements were performed on an inverted microscope (X71, Olympus) in wide-field reflection geometry. Fluorophores were excited through the objective using a supercontinuum pulsed laser (NKT Photonics) coupled to an acousto-optic tunable filter (SuperK Select, NKT Photonics) transmitting a narrow band centered at 620 nm. Fluorescence was detected from ~640 to 700 nm. The repetition rate was 39 MHz. The excitation polarization was set parallel to the axis of the hollow nanocone base. Photoactivation was achieved using a continuous-wave 405 nm laser diode (Oxxius) coupled into the same excitation path.

A high numerical aperture (NA) objective was used to maximize photon collection (oil immersion, NA 1.49, 100×). Fluorescence emission was collected by the same objective and filtered with a long-pass emission filter (BLP01-633R, Semrock), yielding an effective detection band of approximately 640–700 nm. The signal was then split into two paths with a 50:50 beamsplitter and detected by an EMCCD camera (iXon Ultra, Andor) and an eight-pixel SPAD array, respectively. The EMCCD exposure time was set to 30 ms. The SPAD array was connected to a time-correlated single-photon counting (TCSPC) module[62]. Before measurements, the position of each SPAD with respect to the EMCCD field of view (FOV) was calibrated by scanning a 200 nm fluorescent bead in the sample plane and maximizing the signal on each SPAD using a motorized stage.

To achieve single-molecule super-resolved maps, sample drift must be minimized with nanometer precision. We used a piezoelectric nanopositioner (PInano XYZ, Physik Instrumente) controlled by a custom stabilization-loop algorithm. A secondary excitation beam was focused onto a fluorescent bead that served as a fiducial marker. The algorithm maintained the bead position in z (using the measured intensity as a metric) and in x,y (using the localized position) based on EMCCD measurements. The fiducial bead was chosen far from the FOV corresponding to the SPAD array to avoid contaminating the single-molecule signals.

## Single-emitter fluorescence lifetime processing

The principle of the experiment and the smFLIM data correlation are shown in **Fig. 2b**. The SPADs detect single fluorescence photons, and detection events are time-tagged relative to the laser excitation pulses using high-speed electronics (≈1 ps resolution), enabling reconstruction of fluorescence decay curves. To determine the total spontaneous emission decay rate $\Gamma$ (radiative plus non-radiative) for each single emitter, we performed TCSPC to obtain the decay histogram, where $\tau = 1/\Gamma$ is the excited-state lifetime. A repetition rate of 39 MHz was used to allow the emitter to relax to the ground state before the next pulse. The customized TCSPC system and SPAD array provide high detection efficiency (>30% at 650 nm) together with low timing jitter (<60 ps full width at half-maximum, FWHM)[61,62].

By correlating intensity bursts recorded by the SPADs with events simultaneously detected on the EMCCD, we assigned a fluorescence decay rate to each localized emission event with sub-diffraction spatial resolution.

## Single-molecule localization microscopy data analysis

Single-molecule localizations were obtained by fitting each point spread function (PSF) with a 2D Gaussian using the ThunderSTORM[63] plugin in ImageJ[64]. The resulting localizations were further processed with a custom MATLAB script. Single-molecule detections appearing in multiple consecutive frames were merged and then associated with the time-resolved data recorded by the TCSPC. Frames in which multiple detections overlapped with the FOV of a single SPAD were excluded to avoid mixing decay rate information from different molecules.

Decay rates were estimated by fitting each decay histogram with a mono-exponential function convoluted with the instrument response function (IRF), allowing a small temporal shift between the IRF and the measured histogram. A background term was included to account for luminescence and SPAD dark counts. The background level was estimated for each event using time windows in which the molecule was off. Parameter estimation used maximum-likelihood fitting[65], assuming Poisson-distributed photon counts in each time bin.

From the position and decay rate of each correlated single-molecule event, decay rate maps were reconstructed. We represent the data as scatter plots in which each point corresponds to a single measurement, normalized to the average decay rate of molecules on the substrate (glass) surrounded by water in the absence of the nanostructure ($\Gamma_0 \approx 0.3$ ns$^{-1}$). Additional details on smFLIM measurements are provided in Refs. [40,43,44].

## Data overlap for the supercell representation

To increase statistical robustness, we leveraged the periodicity of the HTC lattice to build a "supercell" representation that aggregates localized molecules and their decay rates. The procedure is outlined in **Fig. 3a**. The super-resolved maps in **Fig. 2** were post-processed to determine the center of each nanostructure and define a 2D square lattice used to register all HTC unit cells. Although the lattice period (250 nm) is known with high precision, the lattice origin and orientation were determined for each acquisition (**Fig. 3a**). The SMLM data were then partitioned into 250 × 250 nm unit cells centered on each HTC and superimposed to generate a single supercell containing all correlated molecules within the eight-SPAD array. Because each SPAD is ~1 μm in diameter and therefore contains ~8 complete hollow nanocones, this approach yields ~64× more molecules in the supercell than in a single nanocone.

## Numerical Simulations

All simulations were performed using the Radio Frequency (RF) module of COMSOL Multiphysics[66], which uses the finite element method (FEM) to solve Maxwell's equations in the frequency domain. To match the experimental configuration, we modeled periodic square arrays (unit-cell period 250 nm) of hollow gold nanostructures that are 240 nm high and have a truncated-nanocone geometry with bottom and top outer diameters of 140 nm and 90 nm, respectively. The inner top diameter is 70 nm, corresponding to a top-ring thickness of ~10 nm. Gold permittivity was taken from Johnson and Christy[67].

To represent the substrate supporting the nanostructures, we used a refractive index of 1.5 below the array. The surrounding medium was modeled with a refractive index of 1.33.

## Purcell factor and decay rate calculations

The total decay rate ($\Gamma$) and radiative decay rate ($\Gamma_r$) were computed by modeling an electric point dipole placed at the positions defined in the main text. Dipoles were oriented along the x, y, or z axis (parallel or perpendicular to the nanocone axis, depending on the configuration) and positioned 4 nm from the gold surface of a selected nanostructure.

The total decay rate ($\Gamma$) was derived from the self-interaction Green's function by computing the electric field at the dipole position projected along the dipole moment. The radiative decay rate ($\Gamma_r$) was obtained by integrating the Poynting vector in far-field region, over a closed spherical surface enclosing the system at a distance of ~1.6 μm.

For normalization, $\Gamma$ and $\Gamma_r$ were compared with the corresponding values for an identical dipole in an infinite water medium ($n = 1.33$).

The simulation domain consisted of a 2000 nm × 2000 nm × 2000 nm box surrounded by a 400 nm-thick perfectly matched layer (PML) to absorb outgoing waves. The domain was meshed with tetrahedral elements, with a maximum element size below $\lambda/10$ ($\lambda$ is the wavelength). A finer mesh was applied to the nanostructures. To resolve the near field around the dipole, the source was enclosed in a 6 nm-diameter sphere with refractive index matching water, and the mesh size in this region was <1 nm. Mesh parameters were systematically checked to ensure convergence. High-performance computing resources were required for these calculations.

## Supplementary information

Supplementary information includes: SEM images of the extended lattice and wafer-scale photographs of the 3D hollow plasmonic array (250 nm square lattice); finite-element simulations of modes as a function of wavelength and illumination angle; full-field nano-resolved light–matter interaction maps at oblique illumination before correlation with the SPAD array; angle determination and center-of-mass estimation used to overlap correlated data into a single supercell unit cell; full-field and correlated nano-resolved maps at normal illumination (before and after SPAD correlation); SEM images of the 3D hollow geometry with the cavity filled with polymer; full-field and correlated maps at normal and oblique illumination for polymer-filled cavities (before and after SPAD correlation); simulated total and radiative decay rate enhancements for a dipole attached to the outer wall of the HTC; and calculated response of a point source coupled to the HTC for $P_y$ polarization.

## Acknowledgements

R.M.C.C. thank Rémi Carminati for his valuable discussions. This work has received financial support from the French Agence Nationale de la Recherche via the LABEX WIFI under ANR-10-LABX-24 and ANR-10 IDEX-0001-02 PSL, the ANR SiMpLeLIFe project ANR-17-CE09-0006 and the ANR MIPTIME project ANR-22-CE09-0030-02. R. M. C. C. and R. W. B. acknowledges the funding by the Canada Research Chairs program under award 950-231657, the Natural Sciences and Engineering Research Council of Canada under the Alliance Consortia Quantum grant ALLRP 578468 - 22 and Discovery Grant RGPIN/2017-06880. M. Z-H acknowledges funding from the Gipuzkoa Quantum program's 2024 call of the Department of Economic Promotion and Strategic Projects of the Provincial Council of Gipuzkoa through the grant awarded to the QSEIRA project (2024-QUAN-000011-01).

## Author contributions

R.M.C.C. conceived the project and established collaborations with other institutions. R.M.C.C., I.I., and V.K. led the project development. D.J. and A.S.A. fabricated the hollow truncated nanocone (HTC) arrays. R.M.C.C. performed the smFLIM measurements and analyzed lifetime–position correlations with input from B.V.D., I.I., and V.K. R.M.C.C. developed the fluorescence-labeling method for single emitters on the HTC array. C.C. developed data-filtering functions and tools for statistical analysis and representation of the SMLM results. R.M.C.C. characterized the plasmonic modes in the HTC array, performed the optical measurements, and carried out full-wave numerical simulations. M.Z.H. performed numerical simulations for the Purcell factor. I.I., V.K., and A.S.A. supervised the project. R.M.C.C. wrote the first draft of the manuscript with input from all authors. C.C. created the figures with input from R.M.C.C. All authors commented on the manuscript and contributed to the scientific discussion.

### Competing interests

The authors declare no competing interests.

### Additional information

Correspondence and requests for materials should be addressed to R.M.C.C., I.I. and V. K.

### Abbreviations

LDOS, local density of states; SERS, surface-enhanced Raman scattering; SEM, scanning electron microscopy; HTC, hollow truncated nanocones; TCSPC, time-correlated single-photon counting; LSPR, localized surface plasmon resonance; SMLM, single-molecule localization microscopy; SPAD, single-photon avalanche diode; EMCCD, electron-multiplying charge-coupled device; PSF, point spread function; IRF, instrument response function; FOV, field of view.

# Supplementary material

## Molecule-by-molecule nanoresolved picosecond Purcell-enhanced emission at room-temperature in hollow plasmonic nanocone metamaterials

**R. Margoth Córdova-Castro, [1, 2 *] Dirk Jonker,[3,4]† Clement Cabriel,[1]† Mario Zapata-Herrera,[5] Bart van Dam,[1] Yannick De Wilde,[1] Robert W. Boyd, [2, 6] Arturo Susarrey-Arce,[3,4] Ignacio Izeddin [1 *] and Valentina Krachmalnicoff [1 *]**

[1] Institut Langevin, ESPCI Paris, Université PSL, CNRS, Paris, France.
[2] Department of Physics, University of Ottawa, Ottawa, ON, Canada.
[3] Department of Chemical Engineering, MESA+ Institute, University of Twente, Enschede, The Netherlands.
[4] Mesoscale Chemical Systems, MESA+ Institute, University of Twente, Enschede, The Netherlands.
[5] Donostia International Physics Center (DIPC), San Sebastian, Spain.
[6] University of Rochester, Institute of Optics, Rochester, NY, USA.

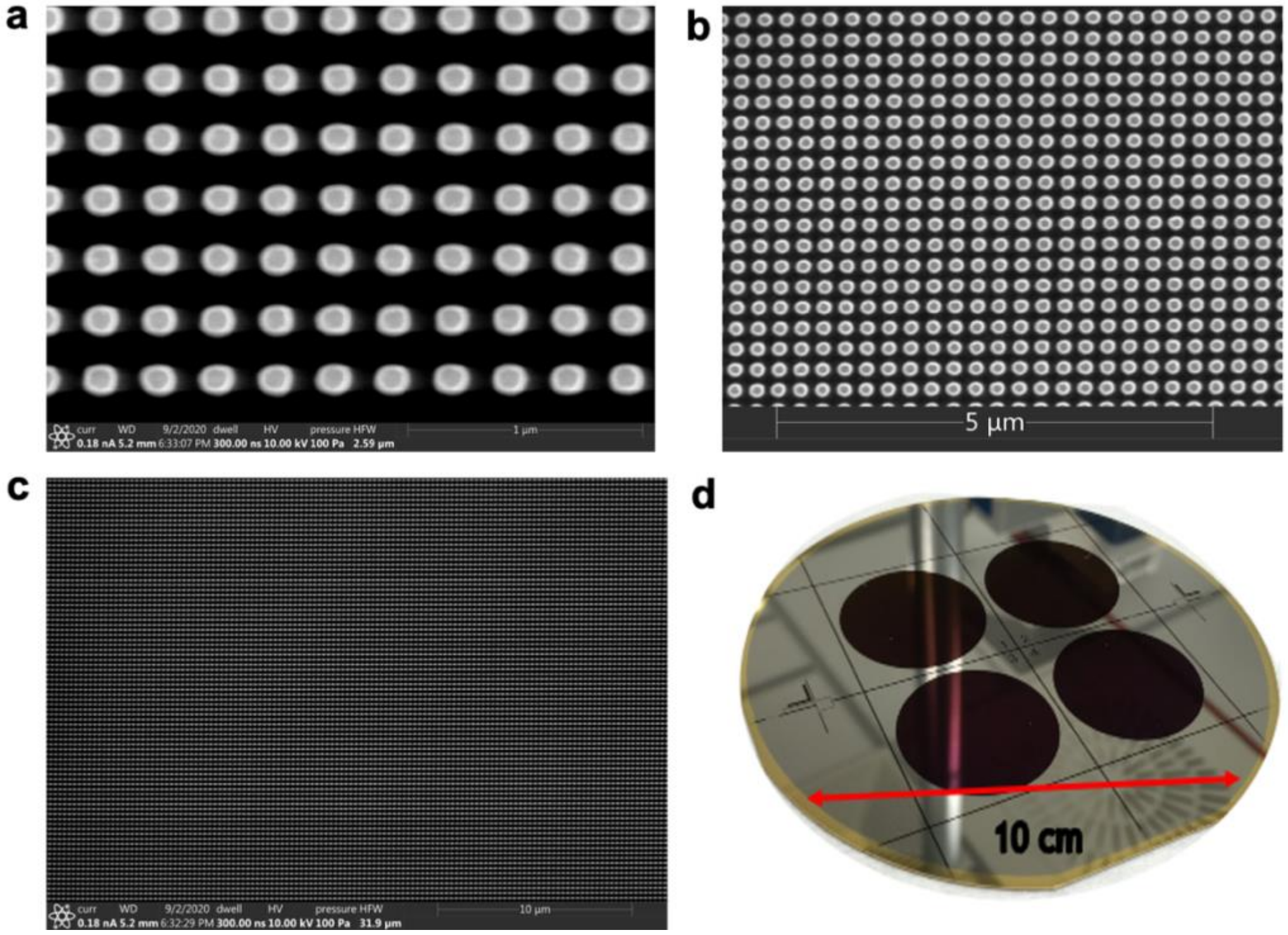

***Fig. S1***: *Wafer-scale 3D hollow plasmonic arrayed in a 250 nm square lattice. Top view Scanning electron microscopy for the Au 3D hollow truncated cones showing the uniform extend of the samples with perfect reproducibility across the wafer scale sample.*

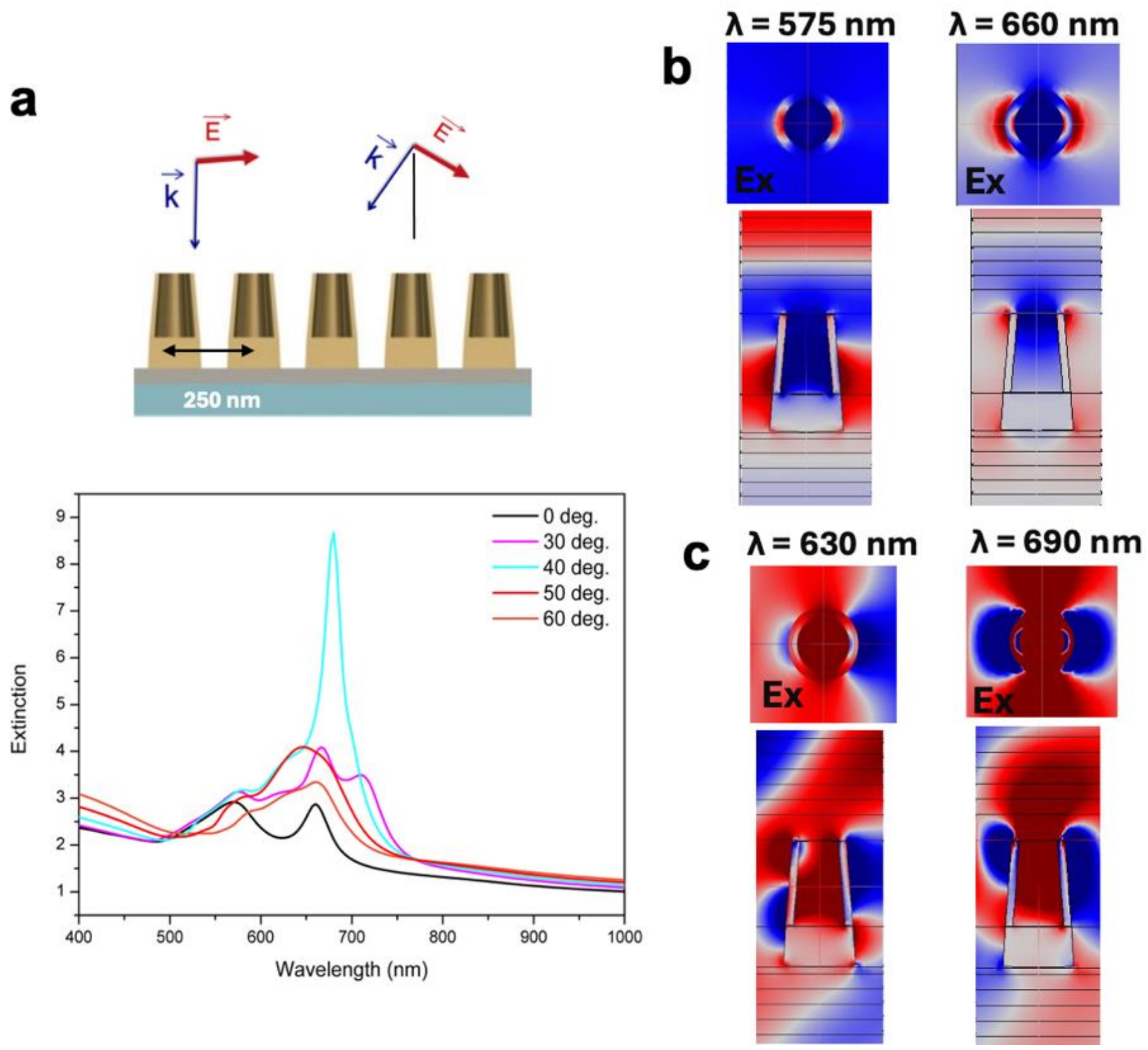

***Fig. S2***: *Finite Element Method calculation of the modes excited as a function of the wavelength and illumination angle. (a) Top: illustration of the geometries corresponding to a normal (left) and an oblique (right) illumination. Bottom: simulation of the extinction spectrum as a function of the illumination angle. (b-c) Normalized simulated near-field electric field maps at different wavelengths for normal (b) and oblique (c) illuminations. The electric field maps are displayed in top view (top) and side view (bottom). These results highlight the diversity of modes that can be excited depending on the illumination conditions.*

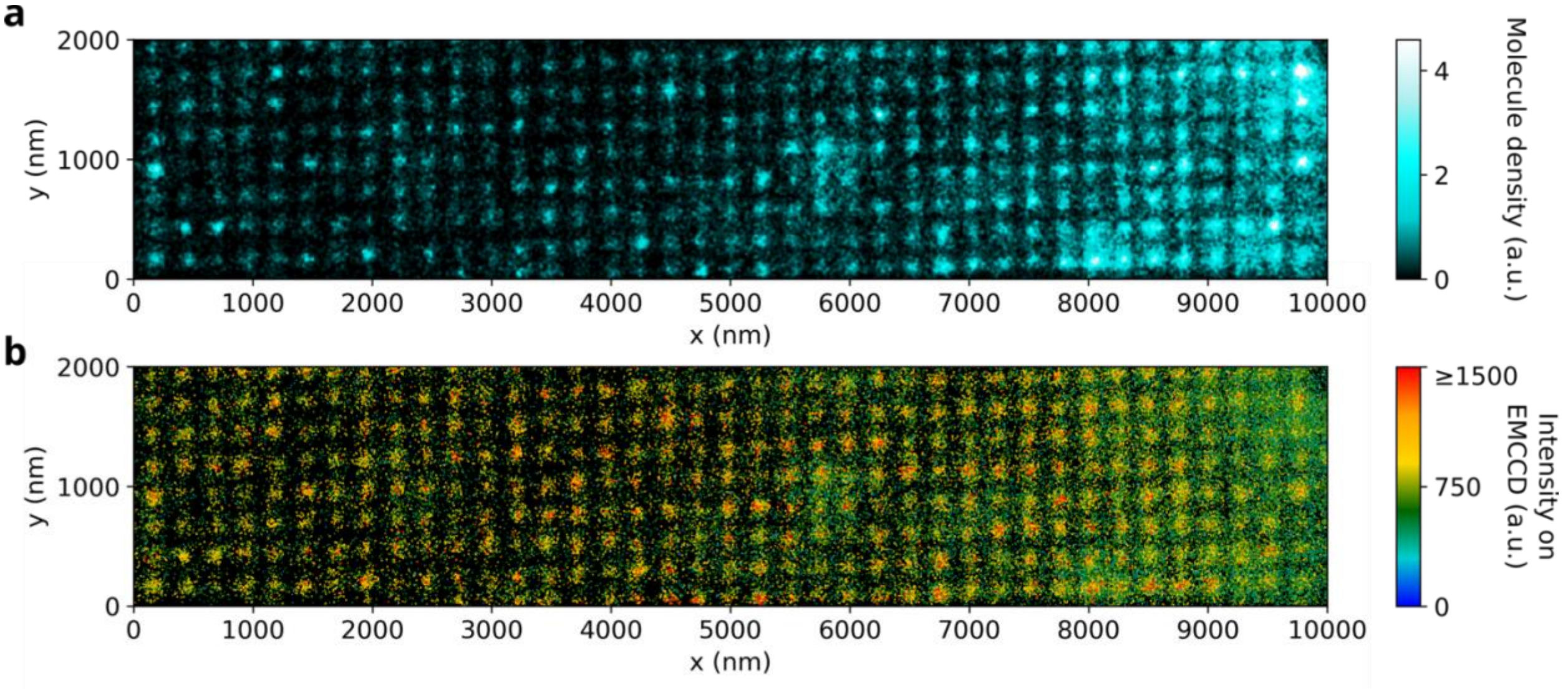

***Fig. S3: Full lattice of the nano-resolved light-matter interaction maps at oblique illumination before correlation with the SPADs array data**. a) Super-resolved map of the density of detected single emitters in an area of 2 µm x 10 µm for an oblique p-polarized pulsed laser illumination at 620 nm. The density of molecules is represented as a pixelized map. (b) Super-resolved map of the molecule-per-molecule collected fluorescence intensity in the EMCCD, represented as a scatter plot. Each point represents one detected molecule, and the color encodes its fluorescence intensity.*

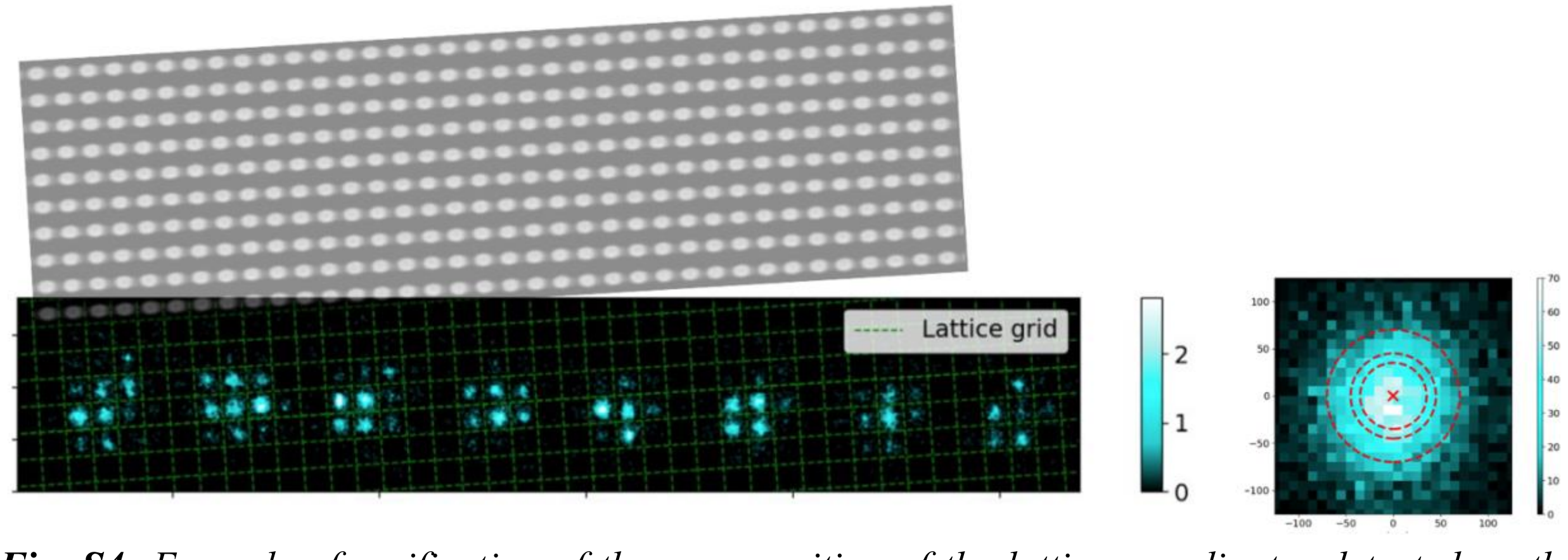

***Fig. S4**: Example of verification of the superposition of the lattice coordinates detected on the single-molecule localization data with the corresponding SEM images (left). The corresponding merged data is displayed in a super unit cell from all correlated data in the 8 single photon detectors (right). Total frame width: 10 µm (left), 250 nm (right).*

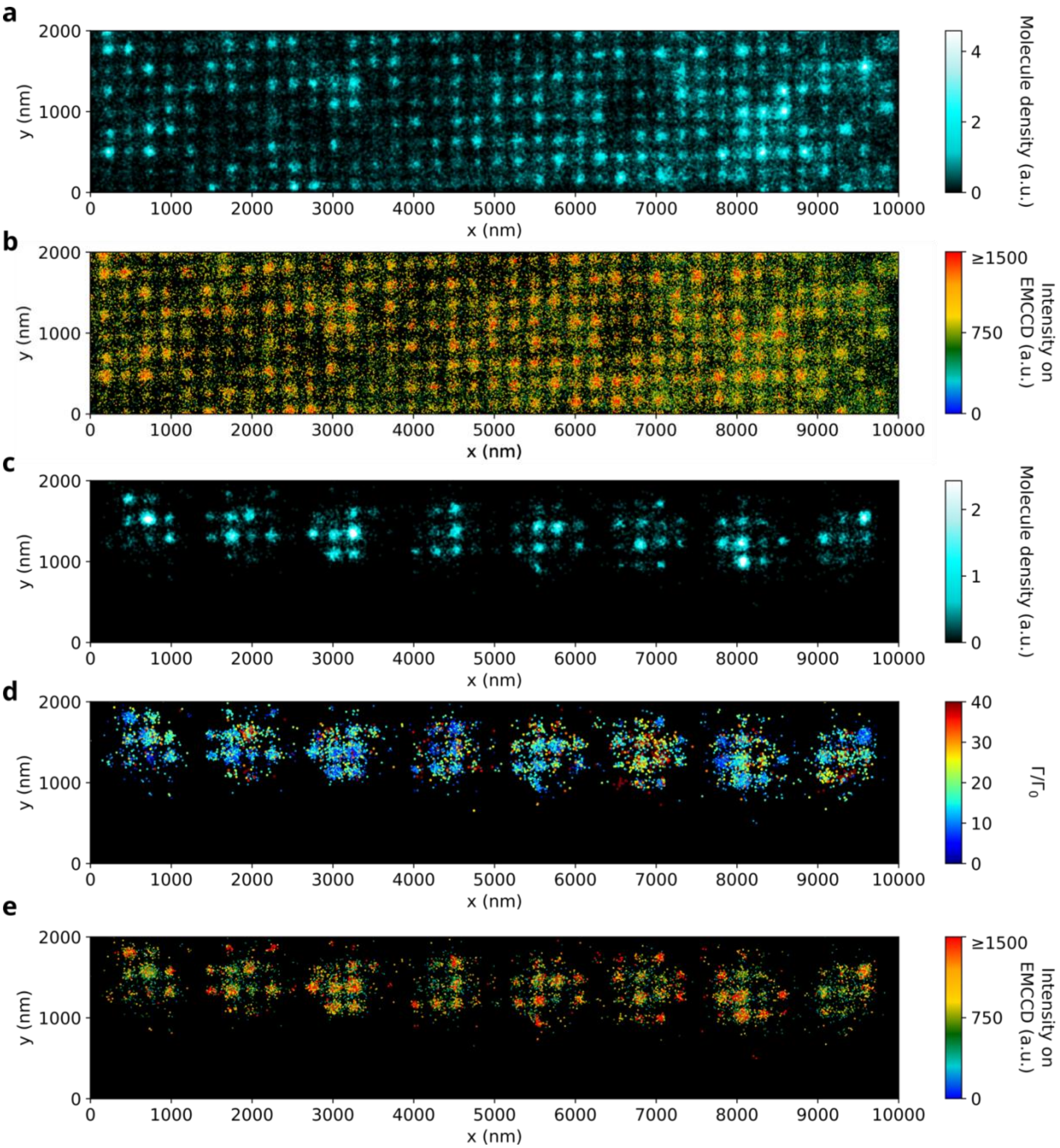

***Fig. S5: Full lattice and correlated lattice of the nano-resolved light-matter interaction maps at normal illumination before and after correlation with the SPADs array data****. a) Super-resolved map of the density of detected single emitters in an area of 2 µm x 10 µm for a normal p-polarized pulsed laser illumination at 620 nm before correlation (full lattice). The density of molecules is represented as a pixelized map. (b) Super-resolved map of the molecule per molecule collected fluorescence intensity in the EMCCD, represented as scatter plot. Each point represents one detected molecule, and the color encodes its fluorescence intensity. c) Super-resolved map of the density of detected single emitters in the same area and under the same illumination conditions after correlation with the SPADs data (correlated lattice). The density of molecules is represented*

*as a pixelized map. (d-e) Super-resolved map of the molecule per molecule collected fluorescence intensity in the EMCCD (d) and the decay rate enhancement (e), represented as scatter plots, after correlation with the SPADs data (correlated lattices). Each point represents one detected molecule, and the color encodes its fluorescence intensity or decay rate enhancement.*

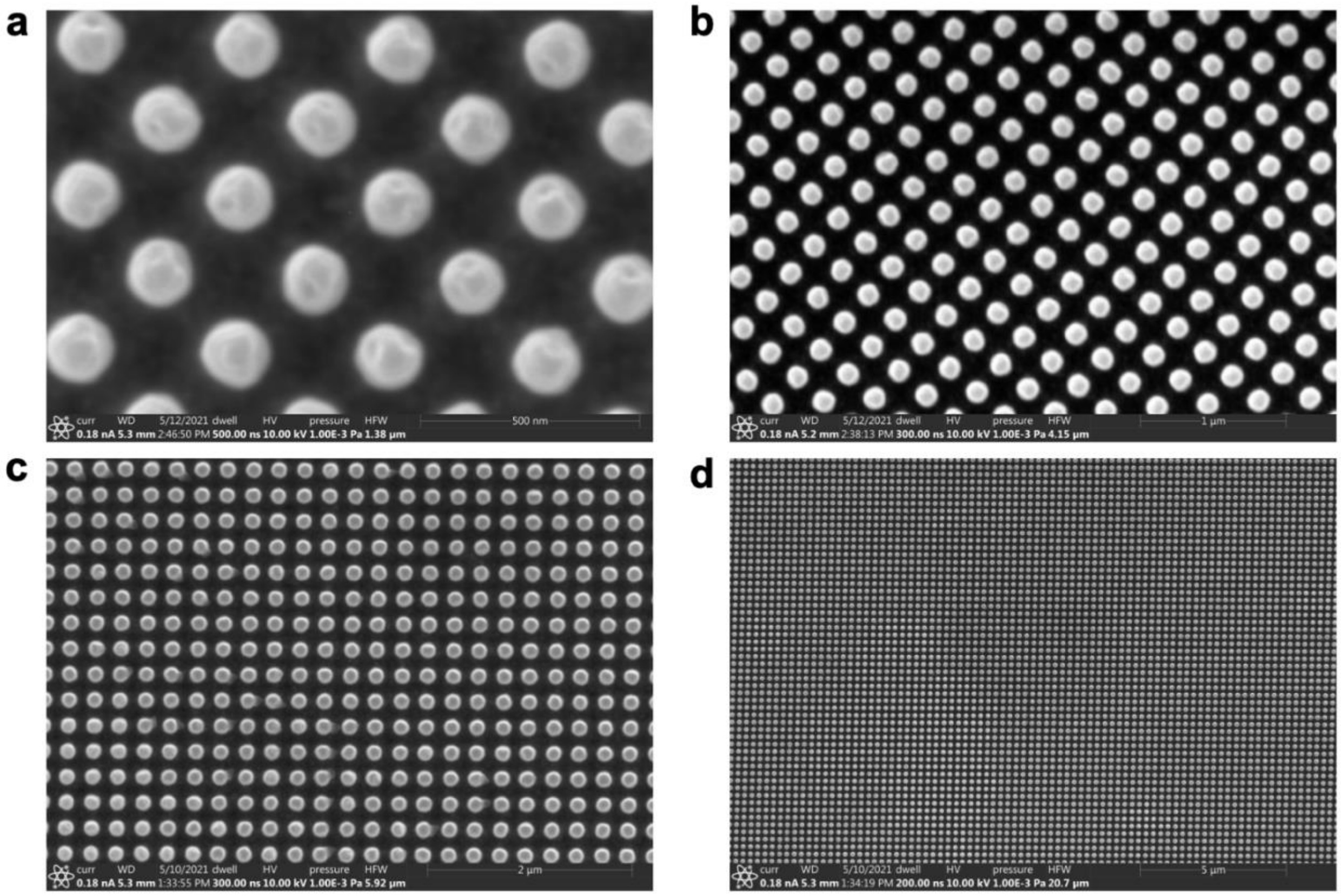

*Fig. S6:* ***3D hollow geometry filled with polymer.*** *Top view Scanning electron microscopy of the Hollow Au truncated cones when filled the cavity with a polymer. A) shows the cover of the full cavity with the polymer. b,c)The extended areas shows that the sample only present the polymer inside the cavity without extra coating out of the cavity pillars. d) Shows the consistence of the sample in large area.*

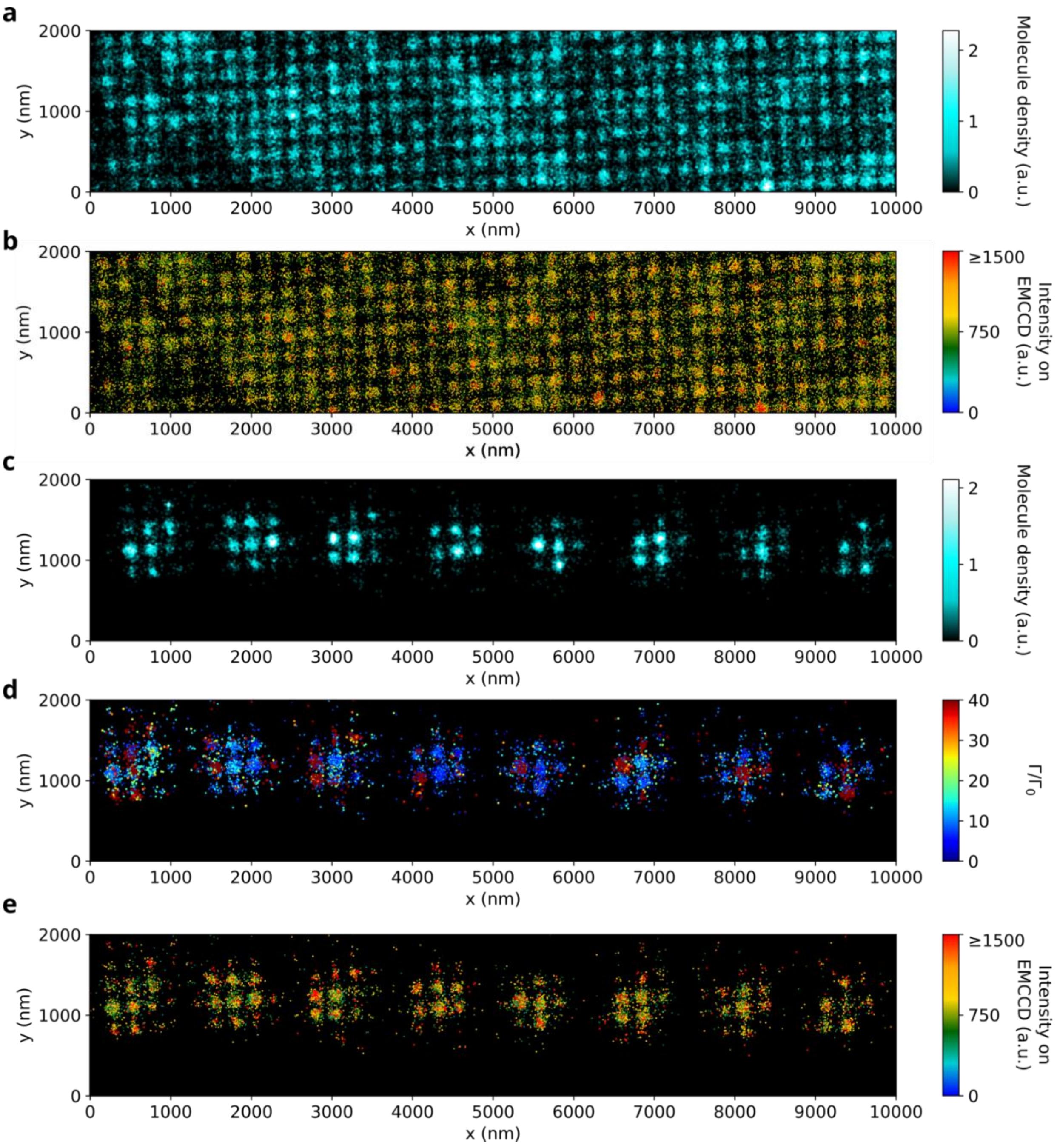

*Fig. S7:* ***Full lattice and correlated lattice of the nano-resolved light-matter interaction maps at normal illumination for a sample with the cavity filled with polymer before and after correlation with the SPADs array data****. a) Super-resolved map of the density of detected single emitters in an area of 2 µm x 10 µm for a normal p-polarized pulsed laser illumination at 620 nm before correlation (full lattice). The density of molecules is represented as a pixelized map. (b) Super-resolved map of the molecule per molecule collected fluorescence intensity in the EMCCD, represented as a scatter plot. Each point represents one detected molecule, and the color encodes its fluorescence intensity. c) Super-resolved map of the density of detected single emitters in the same area and under the same illumination conditions after correlation with the SPADs data*

*(correlated lattice). The density of molecules is represented as a pixelized map. (d-e) Super-resolved map of the molecule per molecule collected fluorescence intensity in the EMCCD (d) and the decay rate enhancement (e), represented as scatter plots, after correlation with the SPADs data (correlated lattices). Each point represents one detected molecule, and the color encodes its fluorescence intensity or decay rate enhancement.*

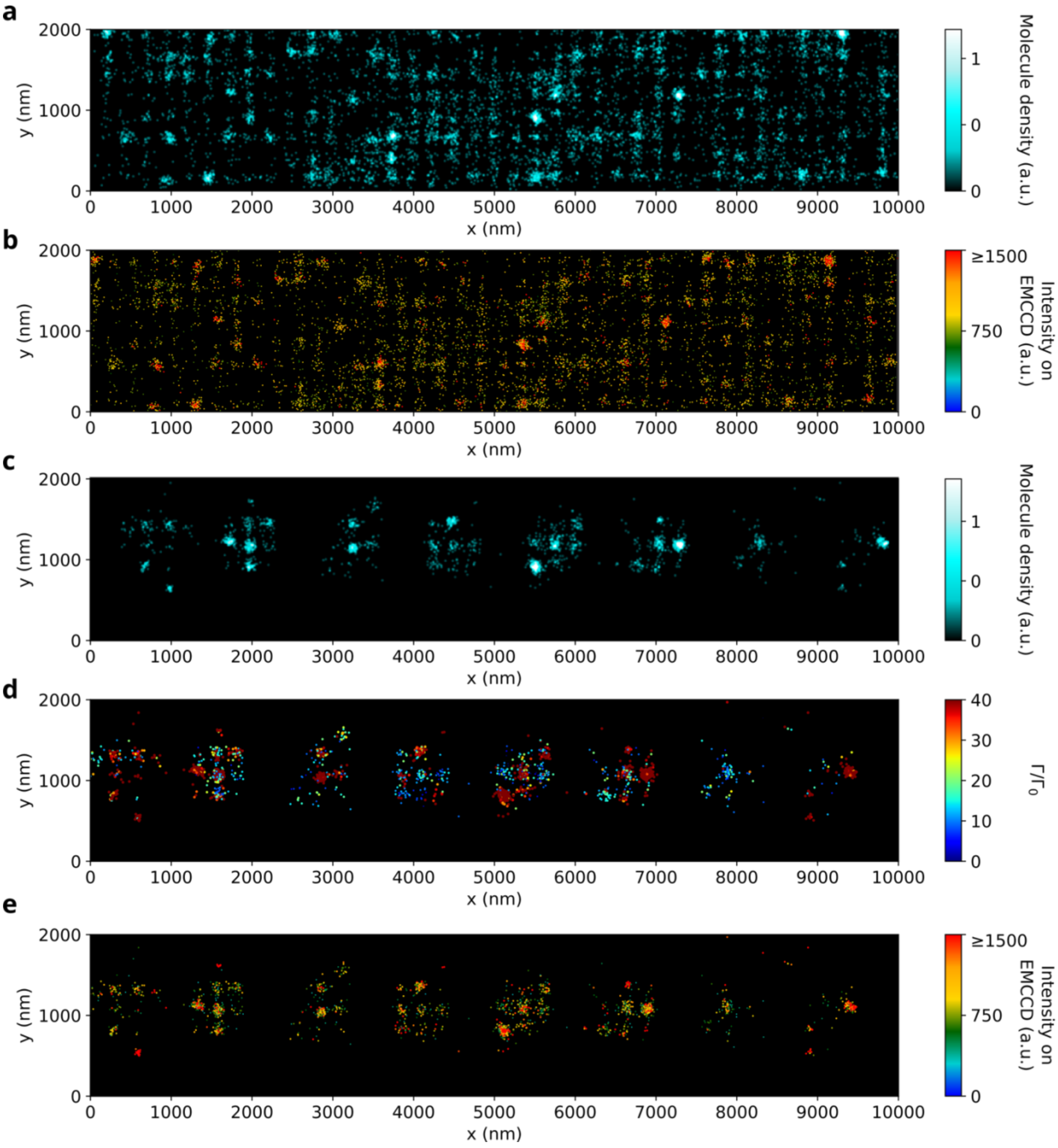

*Fig. S8:* ***Full lattice and correlated lattice of the nano-resolved light-matter interaction maps at oblique illumination for a sample with the cavity filled with polymer before and after correlation with the SPADs array data****. a) Super-resolved map of the density of detected single emitters in an area of 2 µm x 10 µm for a oblique p-polarized pulsed laser illumination at 620 nm before*

*correlation (full lattice). The density of molecules is represented as a pixelized map. (b) Super-resolved map of the molecule per molecule collected fluorescence intensity in the EMCCD, represented as a scatter plot. Each point represents one detected molecule, and the color encodes its fluorescence intensity. c) Super-resolved map of the density of detected single emitters in the same area and under the same illumination conditions after correlation with the SPADs data (correlated lattice). The density of molecules is represented as a pixelized map. (d-e) Super-resolved map of the molecule per molecule collected fluorescence intensity in the EMCCD (d) and the decay rate enhancement (e), represented as scatter plots, after correlation with the SPADs data (correlated lattices). Each point represents one detected molecule, and the color encodes its fluorescence intensity or decay rate enhancement.*

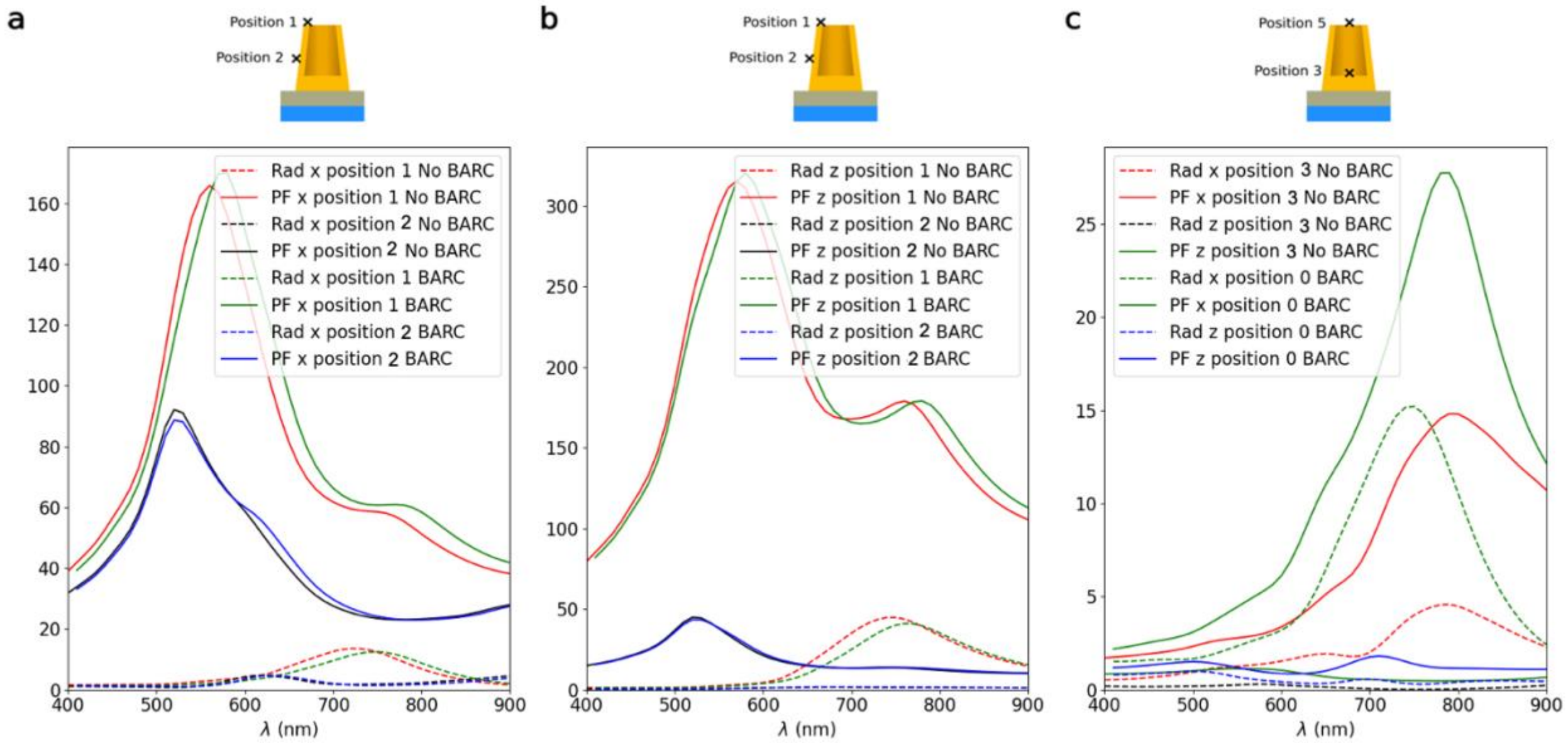

*Fig. S9: Simulated total and radiative decay rate enhancements for a dipole attached to the outer wall of the HTC (separated by 4 nm) at different positions and with different dipole orientations (see the insets), with and without polymer (BARC). Note the different y axes in (a-c). Position 3 is defined only for a non-filled cavity, while position 5 is defined only for a filled cavity. Note that the addition of the polymer changes the decay rate enhancement values only slightly.*

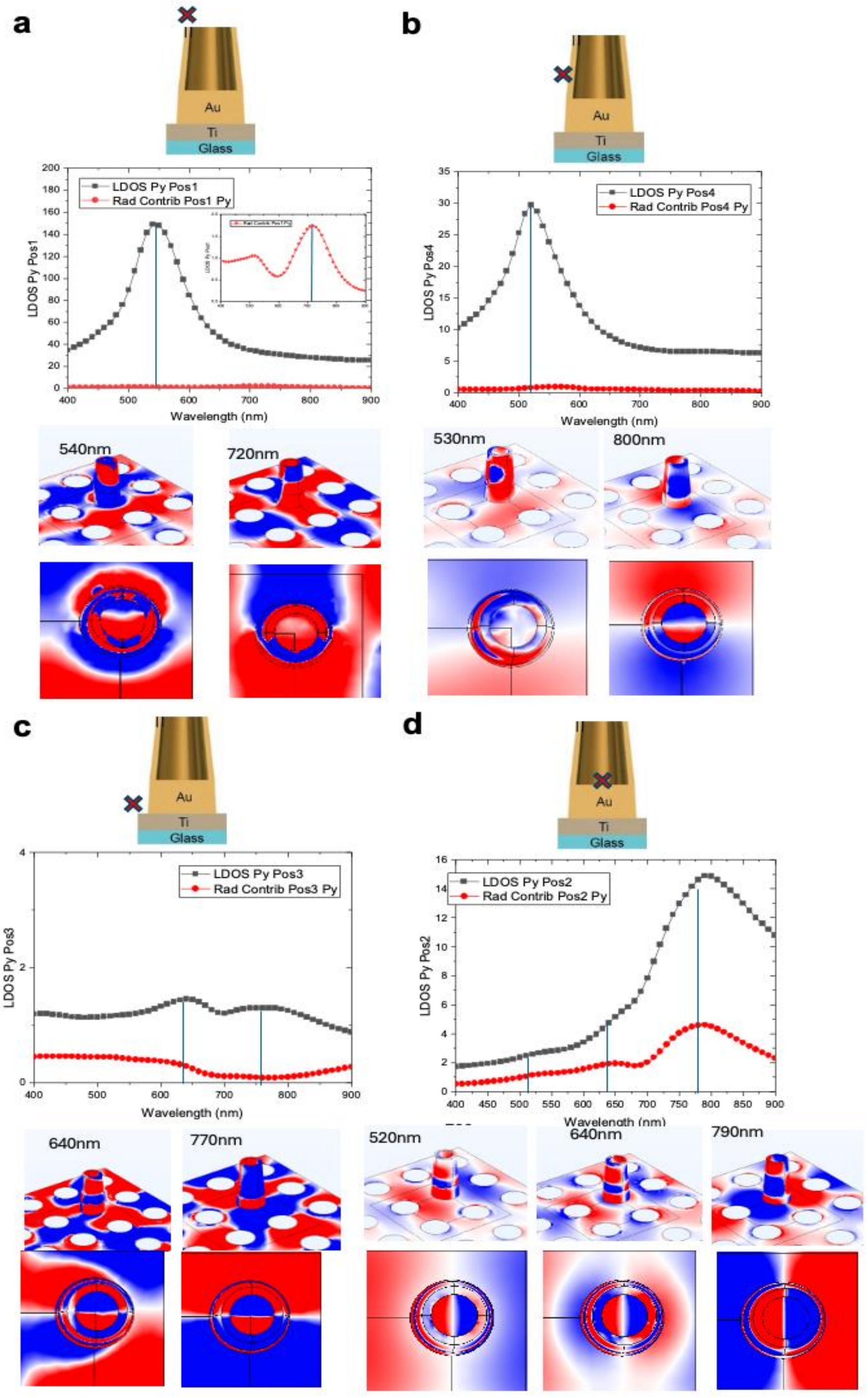

*Fig. S10: Calculated response of a point source coupled to the HTC for Py polarization. a-d) Simulated total and radiative decay rate enhancements for a dipole attached to the outer wall of the HTC (separated by 4 nm) at different positions (see the insets). The decay rate calculated on the top ring (position 1) is higher than the one sensed by the dipole on the outer wall (position 2) or inside the cavity (position 3). Note the different y axes in (a-d). Also shown in the charge distribution for electric dipoles coupled to the nanosystem at different positions and wavelengths. Each disk represents one HTC.*